\documentclass[]{jfm}

\usepackage{graphicx}
\usepackage{newtxtext}
\usepackage{newtxmath}
\usepackage{natbib}
\usepackage{hyperref}
\usepackage{mathtools}
\hypersetup{
    colorlinks = true,
    urlcolor   = blue,
    citecolor  = blue,
}

\newcommand{\RomanNumeralCaps}[1]
\linenumbers
\usepackage{esint}
\usepackage{soul,color,xcolor}
\newcommand{\red}{\textcolor{red}}
\newcommand{\blue}{\textcolor{blue}}

\newcommand{\tabincell}[2]{\begin{tabular}{@{}#1@{}}#2\end{tabular}}

\title[Cavitation bubble interactions with a hemispherical pendant droplet]{Interactions of a collapsing laser-induced cavitation bubble with a hemispherical droplet attached to a rigid boundary}

\shortauthor{Z.~Ren, H.~Han, H.~Zeng, C.~Sun, Y.~Tagawa, Z.~Zuo and S.~Liu}

\author{
Zibo Ren\aff{1}, 
Huan Han\aff{1}, 
Hao Zeng\aff{2},
Chao Sun\aff{2,3,4},
Yoshiyuki Tagawa\aff{5},
Zhigang Zuo\aff{1}\corresp{\email{zhigang200@mail.tsinghua.edu.cn}},
\and
Shuhong Liu\aff{1}\corresp{\email{liushuhong@mail.tsinghua.edu.cn}}}

\affiliation{\aff{1} {State Key Laboratory of Hydro Science and Engineering, and Department of Energy and Power Engineering, Tsinghua University, Beijing 100084, China}

\aff{2}{Center for Combustion Energy, Key Laboratory for Thermal Science and Power Engineering of Ministry of Education, International Joint Laboratory on
Low Carbon Clean Energy Innovation, Department of Energy and Power Engineering, Tsinghua University, Beijing 100084, China}

\aff{3}{Physics of Fluids Group, MESA$^{+}$ Institute and J.~M.~Burgers Centre for Fluid Dynamics, University of Twente, 7500AE Enschede, The Netherlands}

\aff{4}{Department of Engineering Mechanics, School of Aerospace Engineering, Tsinghua University, Beijing 100084, China}

\aff{5}{Department of Mechanical Systems Engineering, Tokyo University of Agriculture and Technology, Tokyo 184-8588, Japan}
}
\begin{document}
\maketitle

\begin{abstract}

We investigate experimentally and theoretically the interactions between a cavitation bubble and a hemispherical pendant oil droplet immersed in water. 
In experiments, the cavitation bubble is generated by a focused laser pulse right below the pendant droplet with well-controlled bubble-wall distances and bubble-droplet size ratios. 
By high-speed imaging, four typical interactions are observed, namely, oil droplet rupture, water droplet entrapment, oil droplet large deformation, and oil droplet mild deformation. 
The bubble jetting at the end of collapse and the migration of the bubble centroid are particularly different in each bubble-droplet interaction.
We propose theoretical models based on the method of images for calculating the Kelvin impulse and the anisotropy parameter which quantitatively reflects the migration of the bubble centroid at the end of the collapse. 
Finally, we explain that a combination of the Weber number and the anisotropy parameter determines the regimes of the bubble-droplet interactions.

\end{abstract}

\section{Introduction}
\label{sec:intro}

Cavitation is a process of the explosive formation and implosive collapse of vaporous bubbles in a liquid. It is a complex phenomenon caused by pressure reduction or energy deposit~\citep{Brennen-17}. Cavitation in a liquid containing particles, droplets, and cells is of great importance in various technological fields. For example, uncontrolled cavitation can cause severe cavitation erosion in hydraulic machinery, especially in a particle-laden liquid~\citep{Karimi86}. The enhancement of cavitation erosion in particle-laden liquids is believed to be related to the interactions between cavitation bubbles and particles, which have achieved many profound understandings in the last two decades~\citep{AroraOhl-40, BorkentArora-41, PoulainGuenoun-326, WuZuo-141,wu2021dynamics,ren2022particulate}. On the other hand, the interactions between well-controlled cavitation and droplets are of particular interest in the fields of ultrasonic cleaning and emulsification. In ultrasonic cleaning, cavitation has been employed to remove dirt, grease, and other contaminants~\citep{MaisonhautePrado-138}. In emulsification, ultrasonic cavitation has been applied to break down large droplets into finer fragments~\citep{MuraCalabria-23, CalifanoCalabria-18, SivaKow-26}. Cavitation has also been employed to provide new ways to deliver a drug into cells or damage the living cells~\citep{KuznetsovaKhanna-25, GacZwaan-32, CoussiosRoy-80, Quinto-SuKuss-34, IinoLi-37, LiYuan-28}. The underlying mechanisms for ultrasonic cleaning and emulsification have been attributed to the complex interactions between droplets (or liquid-liquid interfaces) and cavitation bubbles near boundaries, including the high-speed micro-jetting of collapsing cavitation bubbles, subsequent strong shear flows, shockwave emission, high temperature, or chemical effects~\citep{LiFogler-77, MeroniDjellabi-140}. To identify each effect, past decades have witnessed the study of the interactions between liquid-liquid interfaces and single cavitation bubbles generated by laser pulses~\citep{LauterbornBolle-81} or sparkers. 

Research on the dynamics of bubbles near liquid-liquid interfaces began in the 1980s with experiments on \emph{flat} interfaces, rather than \emph{curved} ones (droplets), and interest in this topic has increased in recent years~\citep{ChahineBovis-291, LiuZhang-146, HanZhang-116}. For example,~\citet{HanZhang-116} investigated the cavitation bubble behaviours near a flat oil-water interface and the subsequent interface jet dynamics with systematic experiments and simulations. They reported that the flow induced by the bubble jetting deformed the liquid-liquid interface and produced the interface jet, which could pinch off and generate daughter droplets. This proves that the microjet at the bubble collapse and the subsequent flows contribute to the formation of emulsified droplets. Although the direction of bubble migration after collapse near a flat liquid-liquid interface can be predicted well by the theory of Kelvin impulse~\citep{BlakeCerone-433, SupponenObreschkow-152}, the bubble behaviours near a curved interface are still not fully investigated. 

Research on the interactions of cavitation bubbles with droplets emerged in the early 2020s.~\citet{YamamotoKomarov-67} performed a three-phase simulation using the volume of fluid method in two comparative systems: a gallium droplet with an air bubble and a silicone oil droplet with an air bubble. Both systems are exposed to ultrasonic waves in a water bath. 
Their simulation results suggest that the physical properties of the droplets, especially the density difference between the phases, are decisive for the direction of the micro-jetting of the cavitation bubble and the subsequent deformation of the droplet. For experiments, ~\citet{OrthaberZevnik-55} generated a single laser-induced cavitation bubble near the interface of a sunflower oil droplet in water and observed from the high-speed photography that the cavitation bubble generates a microjet away from the oil droplet. ~\citet{YamamotoMatsutaka-87} induced cavitation bubbles with ultrasonics near a gallium droplet immersed in water and observed from high-speed photography that the cavitation bubbles collapse and migrate towards the droplet. The bubble jet impacts and ruptures the gallium droplet. 

In previous studies, the droplet was typically more than 10 times larger than the maximum bubble size in radius, minimising the effect of the curved interface on bubble behaviour. ~\citet{AshokeRaman-139} generated single cavitation bubbles in silicone oil near a free-settling water droplet \emph{of comparable radius} and observed the interactions between the cavitation bubble and the water droplet, including deformation, external emulsification and internal emulsification. However, current studies have not considered the interactions between single cavitation bubbles in water and oil droplets of comparable sizes. This is significant because the dispersed phase can be oil droplets in emulsification and ultrasonic cleaning, and their presence may significantly influence cavitation bubble dynamics due to the density difference between the phases and the curvature of the liquid-liquid interface (or bubble-droplet size ratio).  

This study experimentally and theoretically investigates the interactions between a collapsing cavitation bubble and a hemispherical oil droplet attached to a rigid boundary, immersed in water. The ratio of the maximum bubble radius to the droplet size is adjusted. Single cavitation bubbles are generated by focused laser pulses, as detailed in \S\ref{sec:expt}. Typical bubble-droplet interactions are described in \S\ref{sec:exptresults}. A theoretical model is established to predict the displacement of the collapsing bubble in \S\ref{sec: modelling}. Finally, the divisions of different regimes concerning droplet dynamics are proposed in a phase diagram in \S\ref{sec:discussion}.

\section{Experimental set-up}\label{sec:expt}

\begin{figure}
\centerline{\includegraphics[width=130 mm]{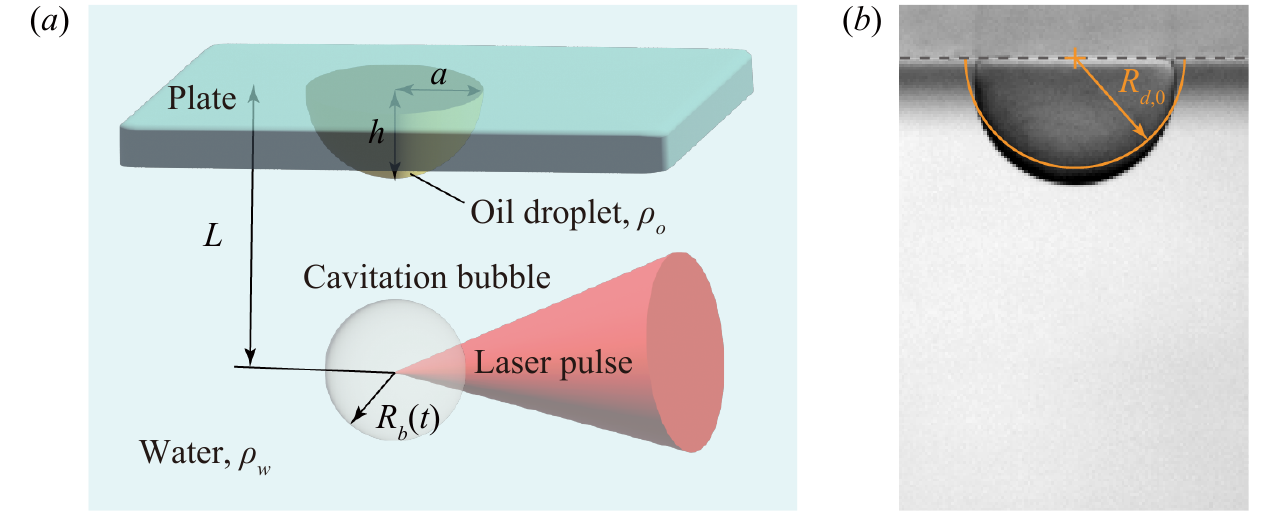}}
\caption{\label{fig:setup} Experimental set-up and notation. (\textit{a}) Schematic of the experimental configuration and notation for pendant droplet-cavitation bubble interactions. (\textit{b}) Definition of the initial equivalent radius of the droplet $R_{d,0}$.}
\end{figure}

A polymethylmethacrylate (PMMA) plate (size $100\,\text{mm}\times\,50\,\text{mm}\times\,10\,\text{mm}$) is fixed horizontally in a quartz chamber filled with degassed and deionised water, see figure~\ref{fig:setup}(\textit{a}).  
A pendant oil droplet is drawn with a disposable syringe and attached to the bottom of the PMMA plate with a long stainless steel needle from outside the chamber~\citep{WangXu-478, WangLi-453}.
To investigate the influences of the density ratio between water (density $\rho_{w}=9.97\,\times\,10^2\,\text{kg}/\text{m}^3$) and oil, we select two kinds of immiscible oil, namely the silicone oil and the kerosene. The silicone oil has a density of $\rho_{o}=(9.60\,\pm\,0.01)\,\times\,10^2\,\text{kg}/\text{m}^3$ and a viscosity of $\mu_{o} = 50\,\text{mPa}\,\text{s}$, while the kerosene has a density of $\rho_{o}=(7.99\,\pm\,0.02)\,\times\,10^2\,\text{kg}/\text{m}^3$ and a viscosity of $\mu_{o} =(1.36\,\pm\,0.01)\,\text{mPa}\,\text{s}$. The densities of both liquids are measured by an electronic densimeter (JHY-120G, Jinheyuan). The viscosity of the kerosene is measured by an Ubbelohde viscometer at $25\,^\circ$C. With the pendant drop method as performed by~\cite{ZengLyu-348}, the surface tension is measured as $(42\,\pm\,4)\,\text{mN/m}$ at the water-silicone oil interface and $(39\,\pm\,4)\,\text{mN/m}$ at the water-kerosene interface. The static shapes of the pendant oil droplets of two kinds are both spontaneously formed as approximate hemispheres (static contact angles $\approx\,90^{\circ}$), with contact radii $a$ and thicknesses $h$, as is shown in figure~\ref{fig:setup}(\textit{a}). The distribution of the sphericity of the oil droplets is detailed in appendix~\ref{sec:app_expt}. The size of the approximate hemispherical oil droplet can be characterised by the initial effective radius $R_{d,0}$ with the same volume as a hemisphere, as shown in figure~\ref{fig:setup}(\textit{b}). We generate pendant silicone oil droplets with $R_{d,0}=1.4$ -- 9.0 mm and pendant kerosene droplets with $R_{d,0}=2.5$ -- 4.2 mm.

Single cavitation bubbles with maximum radii $R_{b,\max}=\,$0.4 -- 3.0~mm are generated by a Q-switched pulsed ruby laser (QSR9, Innolas, with wavelength 694.3~nm, maximum pulse energy 1.5~J, pulse duration 20 -- 30 ns), or by a Q-switched pulsed Nd-YAG laser (LPS-532-L, Changchun New Industries Optoelectronics Technology, wavelength 532 nm, maximum pulse energy 450 mJ, pulse duration 10 ns). The two lasers are used for their different maximum output energy and jitter to produce cavitation bubbles with a wide range of maximum radii. The sphericity of the laser plasma is crucial to the shape of the cavitation bubble~\citep{TagawaYamamoto-369,XuLi-371}. To generate a spherical plasma, the laser beam is expanded 7.5 times before being focused by a convex lens with a focal length of 50 mm, thus forming a convergence angle of about 40 degrees, as performed by~\citet{WuZuo-141, wu2021dynamics}. 

The distance from the cavitation bubble to the PMMA plate $L$ is controlled by positioning the plate with a three-dimensional translation platform. Because of the existence of the boundaries, the bubble is not perfectly spherical, and thus the bubble radius $R_b$ is defined as the effective radius with the same volume as a sphere. We monitor the alignment of the centre of the cavitation bubble on the symmetric axis of the oil droplet with a high-speed camera (FASTCAM Mini UX50, Photron) from the top view and a high-speed camera (v711, Phantom) from the side view. The jitter in the distance from the position of the seeded bubble to the symmetric axis of the oil droplet is controlled within 0.2$\,a$, for details see appendix~\ref{sec:app_expt}. Then a signal generator (9524, Quantum Composers) triggers both the laser and high-speed cameras. The behaviours of the bubble and the droplet are recorded by the high-speed camera from the side view at over $7.9\,\times\,10^4$ frames per second with an exposure time of $1\,\upmu$s. The lens attached to the camera is the same as the ones used in the previous studies~\citep{WuZuo-141,wu2021dynamics,ren2022particulate}.

\section {Overview of the experimental observations}
\label{sec:exptresults}

\subsection{Bubble interactions with a pendant silicone oil droplet in water ($\rho_{o}/\rho_{w}=0.96$)}

\begin{figure}
\centerline{\includegraphics[width=130 mm]{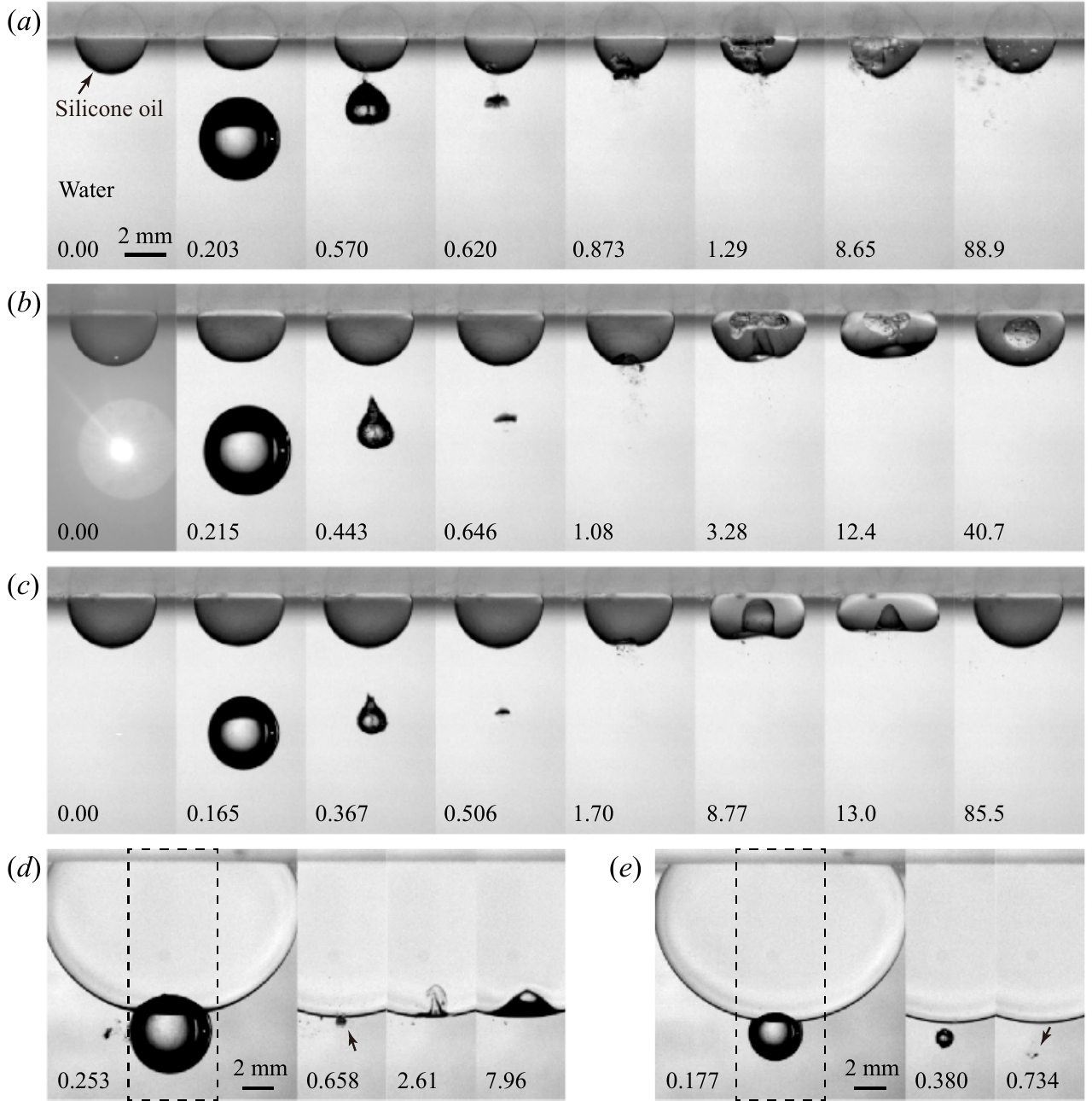}}
\caption{\label{fig:ExpResults_SiO} Snapshots of bubble interactions with pendant silicone oil droplets with density ratio $\rho_{o}/\rho_{w}=0.96$. Cavitation bubbles at collapse migrate towards the droplet (\textit{a}) with rupture of the oil droplet at $L/R_{d,0}=2.74\pm0.08$, $R_{b,\max}/R_{d,0}=1.04\pm0.04$, (\textit{b}) with an emulsified water droplet entrapped inside the oil droplet at $L/R_{d,0}=2.8\pm0.2$, $R_{b,\max}/R_{d,0}=0.87\pm0.07$, (\textit{c}) with large deformation of the droplet at $L/R_{d,0}=2.8\pm0.2$, $R_{b,\max}/R_{d,0}=0.70\pm0.05$, and (\textit{d}) with large deformation of the droplet at $L/R_{d,0}=1.2\pm0.1$, $R_{b,\max}/R_{d,0}=0.29\pm0.02$. In (\textit{e}), the bubble migrates away from the droplet with mild deformation at $L/R_{d,0}=1.2\pm0.1$, $R_{b,\max}/R_{d,0}=0.19\pm0.02$. Photographs in (\textit{a}--\textit{c}) share the same scale bar length of 2 mm, while photographs in (\textit{d}) and (\textit{e}) are zoomed out for better visualisation with their own scale bars. The times are in the units of ms with 0 ms for the laser-plasma generation. The movies are integrated and provided online as supplementary movie 1 at https://doi.org/xxxx.}
\end{figure}

Figure~\ref{fig:ExpResults_SiO} displays experimental observations of four typical responses of the pendant silicone oil droplets induced by bubble behaviours, namely, oil droplet rupture (figure~\ref{fig:ExpResults_SiO}(\textit{a})), water droplet entrapment (figure~\ref{fig:ExpResults_SiO}(\textit{b})), large deformation of the droplet (figure~\ref{fig:ExpResults_SiO}(\textit{c, d})), and mild deformation of the droplet (figure~\ref{fig:ExpResults_SiO}(\textit{e})). For droplets with the same initial effective radius, $R_{d,0}$, different types of interactions are realised by adjusting $L$ and $R_{b,\max}$. Therefore, two dimensionless numbers are proposed, namely, the nondimensional distance from the bubble centre to the plate $L/R_{d,0}$, and the ratio of the bubble maximum radius to the effective droplet radius $R_{b,\max}/R_{d,0}$.

Bubble behaviours are controlled by the compositions of $L/R_{d,0}$ and $R_{b,\max}/R_{d,0}$. For oil droplets with $L/R_{d,0}=2.74\pm0.08$ and $R_{b,\max}/R_{d,0}=1.04\pm0.04$ (see figure~\ref{fig:ExpResults_SiO}(\textit{a})), the cavitation bubble grows to its maximum radius at 0.203 ms and generates an upward jet at collapse, which becomes more pronounced at 0.570 ms during the first rebound of the bubble. Then the bubble jet penetrates the upper interface, forming a vortex ring bubble (0.620 ms). The vortex ring bubble migrates upwards in the water due to its initial inertia and buoyancy and then collides with the bottom of the oil droplet (0.873 ms). The entrained water flows enter the oil droplet, impact the plate, and spread radially. Because of the strength of the vortex ring, the oil droplet is stretched and ruptured (88.9 ms), with dispersive water droplets with a maximum radius of $\approx\,200\,\upmu$m entrapped in the oil droplet.

With similar $L/R_{d,0}$ and smaller $R_{b,\max}/R_{d,0}$, the bubble jetting behaviours are reasonably weakened, see figure~\ref{fig:ExpResults_SiO}(\textit{b}). With $L/R_{d,0}=2.8\pm0.2$ and $R_{b,\max}/R_{d,0}=0.87\pm0.07$, the jet direction of the bubble after the collapse is still upward (0.443 ms), but the initial inertia of the vortex ring bubble decreases, thus leading to a longer time interval for the vortex ring to migrate toward the droplet (1.08 ms). Compared with the case in figure~\ref{fig:ExpResults_SiO}(\textit{a}), this time the kinetic energy of the vortex ring is not high enough to overcome the increase in the surface energy of the droplet. As the kinetic energy dissipates, the entrained water jet pinches off, thus leaving a water droplet with a radius of about 1.7 mm entrapped in the oil droplet. Furthermore, inside the water droplet, multiple oil droplets are distributed with a maximum radius of about $200\,\upmu$m, indicating that O/W/O (oil in water in oil) structures are generated. 

After a period of about several minutes, the internal water droplet in figure~\ref{fig:ExpResults_SiO}(\textit{b}) falls onto the bottom of the oil droplet and merges with the bulk water through the water-oil interface, which is beyond the scope of this work and not discussed in the following sections.

By further decreasing $R_{b,\max}/R_{d,0}$ while maintaining $L/R_{d,0}$ (figure~\ref{fig:ExpResults_SiO}(\textit{c})), the bubble still generates an upward jet at collapse which is more pronounced during bubble rebound (0.367 ms). This time due to the impact of the lifting vortex ring bubble on the droplet, a large and thick inward water column develops from the bottom of the oil droplet (8.77 ms) and induces violent oscillations (13.0 ms) until the droplet recovers to its original state.

The experimental observations above show that the cavitation bubble generates an upward jet at collapse, which is consistent with the
well-known bubble dynamics near a rigid boundary without attached droplets~\citep{LauterbornBolle-81}. Near a flat oil-water interface, a collapsing bubble in water generates a jet away from the interface~\citep{HanZhang-116}, which is not seen in the snapshots shown above. 

To illustrate the effect of the oil-water interface of the droplet on the bubble dynamics, we show two cases in figure~\ref{fig:ExpResults_SiO}(\textit{d}, \textit{e}) where the nondimensional distances are the same ($L/R_{d,0}\approx\,1.2$) with different $R_{b,\max}/R_{d,0}$. In figure~\ref{fig:ExpResults_SiO}(\textit{d}), with $R_{b,\max}/R_{d,0}\approx\,0.3$, the bubble contacts the oil droplet during growth, inducing a local deformation of the droplet (0.253 ms). In this case, the cavitation bubble still migrates upwards after the collapse and evolves into a vortex ring bubble (as marked by the arrow at 0.658 ms). The vortex ring enters the droplet (2.61 ms) and induces only small oscillations of the interface (7.96 ms) until recovery. In figure~\ref{fig:ExpResults_SiO}(\textit{e}), with $R_{b,\max}/R_{d,0}\approx\,0.2$, by contrast, the cavitation bubble migrates away from the oil-water interface, as marked by the arrow at 0.734 ms. This time the subsequent flows induced by the collapsing cavitation bubble are not strong enough to cause the droplet deformation. 

By comparing the observations in figure~\ref{fig:ExpResults_SiO}(\textit{d}, \textit{e}), with $L/R_{d,0}\approx\,1.2$, it is seen that the moving direction of the cavitation bubble is sensitive to $R_{b,\max}/R_{d,0}$. It is to be revealed how the composition of ($L/R_{d,0}$, $R_{b,\max}/R_{d,0}$) determines the bubble centroid migration during the collapse. In the meantime, the small density difference between the silicone oil and the water leads to the bubble jet and motion away from the oil droplet only when $L/R_{d,0}$ is very close to 1.0 and $R_{b,\max}/R_{d,0}$ is smaller than 0.3. This limits our investigation into the repelling phase of the bubble motion, which therefore requires a larger density difference between oil and water. To this end, in the next section, we show experimental observations of bubble interactions with a pendant kerosene droplet. 

\subsection{Bubble interactions with a pendant kerosene droplet in water ($\rho_{o}/\rho_{w}=0.80$)}

\begin{figure}
\centerline{\includegraphics[width=130 mm]{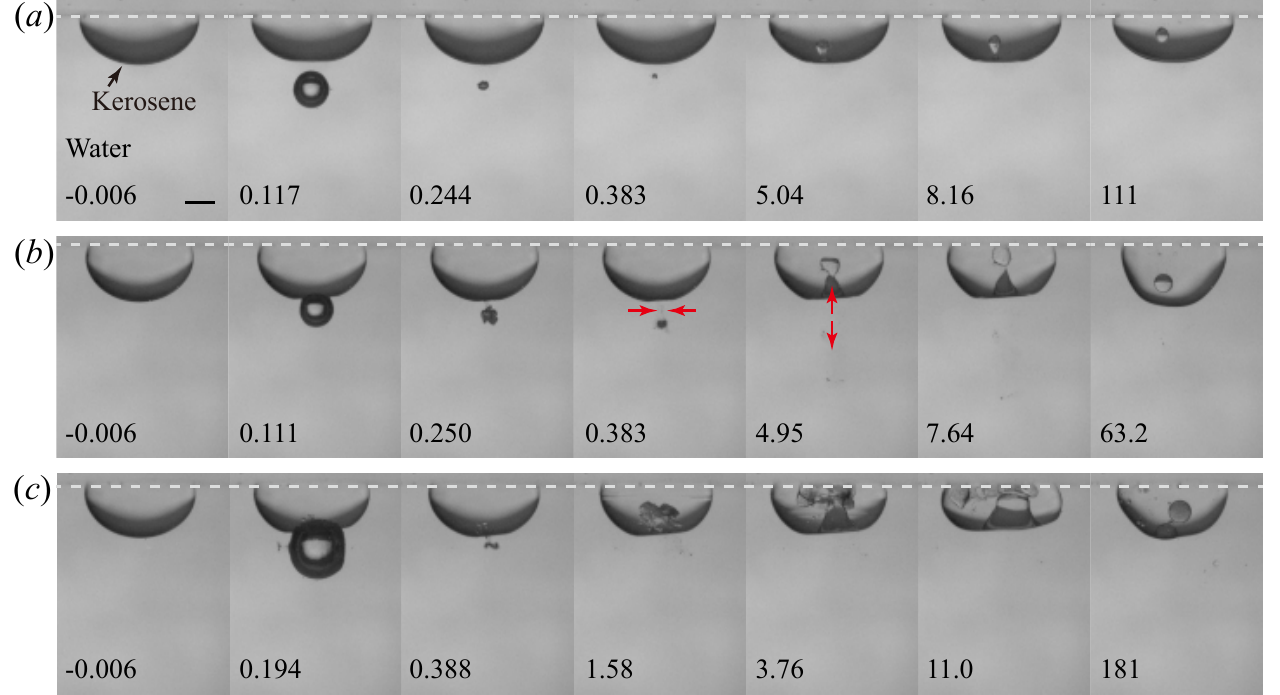}}
\caption{\label{fig:ExpResults_WK} Snapshots of bubble interactions with pendant kerosene droplets with density ratio $\rho_{o}/\rho_{w}=0.80$. (\textit{a}) The bubble at collapse migrates towards the droplet with a water droplet entrapped inside the oil droplet at $L/R_{d,0}=1.3\pm0.1$, $R_{b,\max}/R_{d,0}=0.31\pm0.03$. (\textit{b}) The bubble at collapse migrates away from the droplet inducing an upward focused flow and the entrapment of a water droplet at $L/R_{d,0}=1.14\pm0.03$, $R_{b,\max}/R_{d,0}=0.30\pm0.01$. (\textit{c}) The bubble at collapse migrates towards the droplet leading to the entrapment of emulsified water droplets at $L/R_{d,0}=1.16\pm0.03$, $R_{b,\max}/R_{d,0}=0.47\pm0.02$. Cases in (\textit{a}--\textit{c}) share the same scale bar length of 2 mm. The times are in the units of ms with 0 ms for the laser-plasma generation. The arrows in (\textit{b}) indicate the flow directions. The movies are integrated and provided online as supplementary movie 2 at https://doi.org/xxxx.}
\end{figure}

To look better into the criteria for the migrating direction of the cavitation bubble at collapse, we lower the density ratio of the liquids by replacing silicone oil with kerosene, with the density ratio changing from $\rho_{o}/\rho_{w}=0.96$ to $\rho_{o}/\rho_{w}=0.80$, see figure~\ref{fig:ExpResults_WK}. 

Figure~\ref{fig:ExpResults_WK}(\textit{a}) displays bubble interactions with a pendant kerosene droplet with $L/R_{d,0}=1.3\pm0.1$ and $R_{b,\max}/R_{d,0}=0.31\pm0.03$ when the bubble migrates towards the oil droplet after the collapse (0.244 ms and 0.383 ms). The bubble does not evolve into a vortex ring bubble, but the bubble migration induces a jetting flow of water which enters the oil droplet (5.04 ms). The water column pinches off (8.16 ms) and a single water droplet is entrapped inside the oil droplet. 

In figure~\ref{fig:ExpResults_WK}(\textit{b}), with similar $R_{b,\max}/R_{d,0}$ to figure~\ref{fig:ExpResults_WK}(\textit{a}) and smaller $L/R_{d,0}$, the bubble migrates downwards at collapse (0.250 ms and 0.383 ms). Then the migrating bubble induces a focused radial flow between the bubble and the droplet, which collides (arrows at 0.383 ms) and evolves into axial flows in opposite directions (arrows at 4.95 ms). The upper water flow enters the droplet and pinches off (7.64 ms), leaving a single water droplet entrapped inside the oil droplet. With similar bubble behaviours at collapse, compared with the case for bubble behaviours near a silicone oil droplet (figure~\ref{fig:ExpResults_SiO}(\textit{e})), the focused flow induced by bubble behaviours near a kerosene droplet is much stronger.

Figure~\ref{fig:ExpResults_WK}(\textit{c}) displays observations of bubble-droplet interactions with $L/R_{d,0}$ similar to figure~\ref{fig:ExpResults_WK}(\textit{b}) and larger $R_{b,\max}/R_{d,0}$. The cavitation bubble migrates towards the oil droplet after the collapse and evolves into a vortex ring bubble (0.388 ms) with relatively high kinetic energy, finally leading to O/W/O structures in a similar manner to the case shown in figure~\ref{fig:ExpResults_SiO}(\textit{b}). Comparing figure~\ref{fig:ExpResults_WK}(\textit{b}) and~\ref{fig:ExpResults_WK}(\textit{c}), for pendant kerosene droplets, with $L/R_{d,0}\approx\,1.2$, the direction of bubble migration at collapse is also sensitive to $R_{b,\max}/R_{d,0}$. To predict the bubble centroid displacements at collapse, we establish a quantitative model with details and comparisons with experimental results in \S\ref{sec: modelling}.

\section{Bubble dynamics analysis}
\label{sec: modelling}

\subsection{Theoretical model based on the method of images}

Based on experimental observations, theoretical modelling for predicting the displacement of the collapsing bubble is shown by adopting the idea of the method of images in the potential flow~\citep{Cole-48,BestBlake-212} and the Kelvin impulse~\citep{BlakeCerone-433,BlakeGibson-12,SupponenObreschkow-152}.
The method of images reasonably predicts the behaviour of a collapsing bubble near a rigid plane wall as well as complex walls, such as slot or corner geometries with solid walls~\citep{TagawaPeters-420,MolefePeters-4,AndrewsPeters-267}, an air-water interface~\citep{BlakeGibson-12,KiyamaShimazaki-419}, and an oil-water interface~\citep{BlakeCerone-433,HanZhang-116}.
However, in our problem, the hemispherical shape of an oil droplet attached to the solid wall does not allow us to merely apply the idea of the method of images using a point source or sink.
Here we refer to the theory proposed by \citet{Weiss-83} and expand the application of the method of images described as follows.

\begin{figure}
\centerline{\includegraphics[width=130 mm]{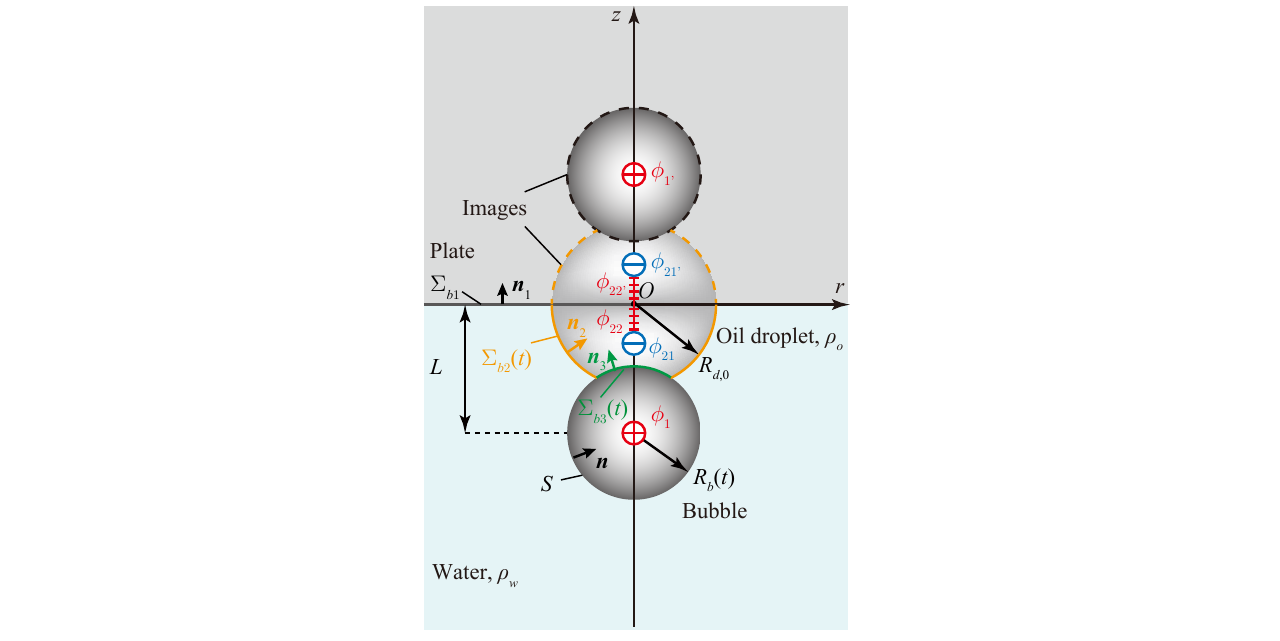}}
\caption{\label{fig:Model} Theoretical model based on the method of images. The velocity potentials are detailed in table~\ref{Table1}. $\phi_1$ and $\phi_{1'}$ represent a point source and its image (red circled plus $\red{\oplus}$). $\phi_{21}$ and $\phi_{21'}$ represent a point sink and its image (blue circled minus $\blue{\ominus}$). $\phi_{22}$ and $\phi_{22'}$ represent uniformly distributed line sources and images (red plus $\red{+}$).}
\end{figure}

As shown in figure~\ref{fig:Model}, the plate is regarded as a rigid boundary with its lower surface set at the plane of $z=0$ in the cylindrical coordinates ($r,z$). Here, because the system is cylindrically symmetric, the circumferential coordinate is omitted. The whole system is immersed in water with density $\rho_{w}$. A pendant oil droplet with density $\rho_{o}$ is assumed to be in a hemispherical shape with radius $R_{d,0}$. The origin $O$ of the coordinate system is placed at the centre of the circular contact line of the droplet. A \emph{spherical} cavitation bubble is generated on the $z$ axis with radius $R_{b}(t)$ varying with time $t$. The coordinate of the bubble centre is set at $z_{b}=-L$. The solid-water interface is denoted by ${\upSigma}_{b1}$, the droplet-water interface by ${\upSigma}_{b2}$, and the droplet-bubble interface by ${\upSigma}_{b3}$ when the bubble and the oil droplet contact. The solid-water interface $\Sigma_{b1}$ does not include the region where the droplet is attached to the solid boundary. In correspondence with our experimental observations shown in figure~\ref{fig:ExpResults_SiO} and figure~\ref{fig:ExpResults_WK}, the droplet-water interface ${\upSigma}_{b2}$ maintains the hemispherical shape as the initial state during the first growth and collapse of the cavitation bubble. We should note that when the bubble and the droplet contact both boundaries ($\Sigma_{b2}$ and $\Sigma_{b3}$) are varying with time.

On the timescale of the lifetime of the cavitation bubble, both the oil droplet and the water can be seen as incompressible liquids~\citep{HanZhang-116}. The Reynolds number related to cavitation bubble dynamics is defined as $\Rey_{b}=R_{b,\max}\sqrt{\upDelta p/\rho_{w}}/\nu_{w}=4\,\times\,10^3$ -- $3\,\times\,10^4\,\gg\,1$, with $R_{b,\max}=0.4$ -- 3.0 mm, the pressure difference driving bubble collapse $\upDelta p=p_{\infty}-p_{v}$, the pressure in static water $p_{\infty}=1.01\times\,10^5\,\text{Pa}$, the vapour pressure in the bubble $p_{v}=2.3\times\,10^3\,{\text{Pa}}$,  $\rho_{w}\approx\,1\,\times\,10^3\,\text{kg}/\text{m}^3$, and the kinetic viscosity of water $\nu_{w}=1\,\times\,10^{-6}\,\text{m}^2/\text{s}$. Therefore the viscosity can be neglected during this stage. Under the assumptions of incompressible, inviscid and irrotational fluids, we formulate a potential flow model to calculate the velocity field $\boldsymbol{v}$ with the method of images, as illustrated in figure~\ref{fig:Model}. 

The cavitation bubble is simulated by a point source (No.~1) with strength $Q(t)=4\upi {R_{b}}^2\dot{R}_{b}$, and thus its velocity potential at any point ($r,z$) in the flow field reads 
\begin{equation}
  \phi_1=-\frac{Q}{4\,\upi\,\lVert(r,z)-(0,z_{b})\rVert}.
\end{equation}

With the method of images, the boundary conditions for the system must be fulfilled at the same time. Referring to the model for bubble dynamics near a flat liquid-liquid interface~\citep{BlakeCerone-433}, the boundary condition at the droplet-water interface $\upSigma_{b2}$ reads,
\begin{equation}
    \rho_{w}\,\Phi_1=\rho_{o}\,\Phi_2,
\label{eq:BC}
\end{equation}
with $\Phi_1$ being the superposed velocity potential in water and $\Phi_2$ in droplet. This condition is called a linearised dynamic boundary condition, considering the force balance across the interface and ignoring the nonlinear effects on the interface (e.g., viscous forces, capillary waves, etc.).

Meanwhile, the fluid velocity normal to the droplet-water interface $\upSigma_{b2}$ should be continuous, leading to
\begin{equation}  \nabla\Phi_1\cdot\,\boldsymbol{n}_2=\nabla\Phi_2\cdot\,\boldsymbol{n}_2,
  \label{eq:BC22}
\end{equation}
with $\boldsymbol{n}_2$ the unit normal vector directing from the water to the droplet.

The boundary condition at the rigid boundary $\upSigma_{b1}$ with zero normal velocity reads,
\begin{equation}    \nabla\Phi_1\cdot\,\boldsymbol{n}_1=\nabla\Phi_2\cdot\,\boldsymbol{n}_1=0,
\label{eq:BC1}
\end{equation}
where $\boldsymbol{n}_1$ is the unit normal vector directing from the water to the rigid boundary.

\begin{table}
\begin{center}
    \def~{\hphantom{0}}
\begin{tabular}{c*{5}c}
No.~$i$ & Basic solution  &  Location   &  Strength    & Velocity potential $\phi_i$  \\
\\
1 & Point source    &  ($0,z_{b}$)    &   $Q$     & $-\frac{Q}{4\,\upi\,\lVert(r,z)-(0,z_{b})\rVert}$ \\
\\
$1^{\prime}$ & Point source   &  ($0,-z_{b}$)    &   $Q$     & $-\frac{Q}{4\,\upi\,\lVert(r,z)-(0,-z_{b})\rVert}$ \\
\\
21 & Point sink    &  ($0,z_{2}$)    &   $-Q\,R_{d,0}/L$     & $\frac{Q\,R_{d,0}\,/\,L}{4\,\upi\,\lVert(r,z)-(0,z_{2})\rVert}$ \\
\\
$21^{\prime}$ & Point sink    &  ($0,-z_{2}$)    &   $-Q\,R_{d,0}/L$     & $\frac{Q\,R_{d,0}\,/\,L}{4\,\upi\,\lVert(r,z)-(0,-z_{2})\rVert}$ \\
\\
22 & \tabincell{c}{Uniformly distributed\\ line sources}    &  \tabincell{c}{From ($0,0$) \\to ($0,z_{2}$)}    &   $Q/R_{d,0}$     & $-\frac{Q}{4\,\upi\,R_{d,0}}\int_0^{\frac{R_{d,0}^2}{L}}\frac{dl}{\lVert(r,z)-(0,-l)\rVert}$ \\
\\
$22^{\prime}$ & \tabincell{c}{Uniformly distributed\\ line sources}    &  \tabincell{c}{From ($0,0$) \\to ($0,-z_{2}$)}    &   $Q/R_{d,0}$     & $-\frac{Q}{4\,\upi\,R_{d,0}}\int_0^{\frac{R_{d,0}^2}{L}}\frac{dl}{\lVert(r,z)-(0,l)\rVert}$\\
\\
\end{tabular}
\caption{Velocity potentials of the basic solutions in the theoretical model, with $z_{b}=-L$ and $z_2=-{R_{d,0}}^2/L$.}
\label{Table1}
\end{center}
\end{table}

To fulfil all the boundary conditions (\ref{eq:BC}) -- (\ref{eq:BC1}) at the same time, we adapt the Weiss sphere model~\citep{Weiss-83} which was proposed for hydrodynamic images in a rigid sphere immersed in arbitrary potential flows. In our model for the flow outside an oil droplet, we convert the signs of the hydrodynamic images in the Weiss sphere model, as shown in figure~\ref{fig:Model}. A point sink (No.~21) and a set of uniformly distributed line sources (No.~22) are placed inside the oil droplet. The coordinate of the point sink (No.~21) is ($0,-{R_{d,0}}^2/L$) with strength $-Q\,R_{d,0}/L$ and velocity potential $\phi_{21}$. The detailed formula is referred to in table~\ref{Table1}. The uniformly distributed line sources (No.~22) extend from the origin (0,0) to the point sink (No.~21), with line density $Q/R_{d,0}$ and velocity potential $\phi_{22}$. Then a mirror point source (No.~1'), a mirror point sink (No.~21'), and a set of mirror line sources (No.~22') are placed symmetrically about the rigid boundary, with their locations, strengths, and velocity potentials shown in table~\ref{Table1}.

Next, the superposed velocity potential in water at ($r,z$) reads 
\begin{equation}
\Phi_1=\phi_1+\phi_{1'}+\frac{\rho_{w}-\rho_{o}}{\rho_{w}+\rho_{o}}\left[\,\phi_{21}+\phi_{21'}+\phi_{22}+\phi_{22'}+F(r,z)\right],
\label{eq:Phi11}
\end{equation}
while in oil the superposed velocity potential reads
\begin{equation}
\Phi_2=\frac{2\,\rho_{w}}{\rho_{w}+\rho_{o}}\,\left(\phi_1+\phi_{1'}\right),
\label{eq:Phi12}
\end{equation}
by referring to~\citet{BlakeCerone-433}. In (\ref{eq:Phi11}), $F(r,z)$ is an additional function to be determined for the fulfilment of the boundary conditions, which has the same unit as the velocity potential. Substituting (\ref{eq:Phi11}) and (\ref{eq:Phi12}) into (\ref{eq:BC}) -- (\ref{eq:BC1}), we get three equations as follows.

On the solid-water interface ${\upSigma}_{b1}$, we obtain
\begin{equation}
  \frac{\partial F(r,z)}{\partial z}\Bigg|_{{\upSigma}_{b1}}=0.
  \label{eq:Frz}
\end{equation}

On the droplet-water interface ${\upSigma}_{b2}$, we obtain
\begin{equation}
\left[\phi_1+\phi_{1'}+\,\phi_{21}+\phi_{21'}+\phi_{22}+\phi_{22'}+F(r,z)\right]\big|_{{\upSigma}_{b2}}=0,
\end{equation}
and
\begin{equation}
\nabla\left(\phi_{1}+\phi_{1'}\right)\big|_{{\upSigma}_{b2}}\cdot\,\boldsymbol{n}_2=\nabla\left[\phi_{21}+\phi_{21'}+\phi_{22}+\phi_{22'}+F(r,z)\right]\big|_{{\upSigma}_{b2}}\cdot\,\boldsymbol{n}_2.
\end{equation}

With the design of the hydrodynamic images, the velocity potentials are known to satisfy
\begin{equation}  \left(\phi_1+\phi_{21}\right)\big|_{{\upSigma}_{b2}}=\left(\phi_{1'}+\phi_{21'}\right)\big|_{{\upSigma}_{b2}}=0,
\end{equation}
and
\begin{equation}
\nabla\left(\phi_{1}+\phi_{1'}\right)\big|_{{\upSigma}_{b2}}\cdot\,\boldsymbol{n}_2=\nabla\left[\phi_{21}+\phi_{21'}+\phi_{22}+\phi_{22'}\right]\big|_{{\upSigma}_{b2}}\cdot\,\boldsymbol{n}_2.
\end{equation}

Thus, besides (\ref{eq:Frz}), the additional function $F(r,z)$ satisfies
\begin{equation}   \left[\phi_{22}+\phi_{22'}+F(r,z)\right]\big|_{{\upSigma}_{b2}}=0,
   \label{eq:Frz2}
\end{equation}
and
\begin{equation}
  \nabla F(r,z)\big|_{{\upSigma}_{b2}}\cdot\,\boldsymbol{n}_2=0.
  \label{eq:Frz3}
\end{equation}

The derivation of the additional function $F(r,z)$ is detailed as follows. The relation (\ref{eq:Frz3}) indicates the condition of zero component velocity normal to the droplet-water interface ${\upSigma}_{b2}$. When we convert the cylindrical coordinates ($r,z$) to the polar coordinates ($R,\varphi$) in the same plane, as shown in the inset of figure~\ref{fig:model_verify}(\textit{a}), the relation (\ref{eq:Frz3}) can be rewritten as
\begin{equation}
\left(\frac{\partial F}{\partial R}\right)\bigg|_{{\upSigma}_{b2}}=0.
\label{eq:FR}
\end{equation}
This indicates that $F(R,\varphi)$ is only a function of $\varphi$. Thus we select one of the forms as
\begin{equation}  F(\varphi)=a_0+a_2\,\cos^2{\varphi}+a_4\,\cos^4{\varphi}+a_6\,\cos^6{\varphi}+a_8\,\cos^8{\varphi},
\end{equation}
where the coefficients $a_0$, $a_2$, $a_4$, $a_6$, and $a_8$ are determined by the least square fitting with (\ref{eq:Frz2}).

When the bubble and the droplet contact, on the boundary of $\Sigma_{b3}$, the boundary condition is simplified as the velocity equal to $\nabla\Phi_2$ because of the inward deformation of the droplet. When the bubble does not contact with the droplet, only $\Sigma_{b1}$ and $\Sigma_{b2}$ exist. Our method is verified in detail in appendix~\ref{sec:app_image}.

\subsection{Calculation of the Kelvin impulse}

The Kelvin impulse $\boldsymbol{I}_{S}$ is often used for the quantitative judgement of the centroid migration of a cavitation bubble~\citep{SupponenObreschkow-152}. Near a single boundary (e.g., rigid boundary, free surface, liquid-liquid interface, etc.), the Kelvin impulse $\boldsymbol{I}_{S}$ is defined as the closed-loop integral of the velocity potential at the bubble interface $S$, i.e., $\boldsymbol{I}_S=\rho_{w}\,\varoiint_{S}\Phi\,\boldsymbol{n}_s\,\text{d}A$, with $\Phi$ the velocity potential and $\boldsymbol{n}_s$ the unit normal vector directing from the water to the interior of the bubble. According to \citet{BlakeCerone-433}, the Kelvin impulse can be written as
\begin{equation}
\boldsymbol{I}_S=\rho_{w}\,\int_0^t\,\int_{\upSigma}\left[\frac{1}{2}\lVert\nabla\Phi\rVert^2\boldsymbol{n}-\frac{\partial\Phi}{\partial\,n}\nabla\Phi\right]\text{d}A\text{d}t,
\label{eq:Is}
\end{equation}
where ${\partial\Phi}/{\partial\,n}=\nabla\Phi\cdot\boldsymbol{n}$ denotes the normal velocity to the boundary $\upSigma$ with $\boldsymbol{n}$ the unit normal vector directing from the liquid to the boundary. 

In our problem, the system contains three boundaries, which indicates that the Kelvin impulse consists of three parts, namely, the contributions from the solid-water interface $\boldsymbol{I}_{b1}$, from the droplet-water interface $\boldsymbol{I}_{b2}$, and from the droplet-bubble interface $\boldsymbol{I}_{b3}$ when the bubble and the droplet contact. 

For boundaries $\upSigma_{b1}$ and $\upSigma_{b2}$, the velocity potential is $\Phi_1$ and the fluid density is $\rho_{w}$, while for boundary $\upSigma_{b3}$ the velocity potential is $\Phi_2$ and the fluid density is $\rho_{o}$. Therefore the three contributions of the Kelvin impulse read
\begin{equation}
\begin{dcases}
\boldsymbol{I}_{b1}=\rho_{w}\,\int_0^t\,\int_{{\upSigma}_{b1}}\left[\frac{1}{2}\lVert\nabla\Phi_1\rVert^2\boldsymbol{n}_1-\frac{\partial\Phi_1}{\partial\,n_1}\nabla\Phi_1\right]\text{d}A\text{d}t,\\
\boldsymbol{I}_{b2}=\rho_{w}\,\int_0^t\,\int_{{\upSigma}_{b2}(t)}\left[\frac{1}{2}\lVert\nabla\Phi_1\rVert^2\boldsymbol{n}_2-\frac{\partial\Phi_1}{\partial\,n_2}\nabla\Phi_1\right]\text{d}A\text{d}t,\\
\boldsymbol{I}_{b3}=\rho_{o}\,\int_0^t\,\int_{{\upSigma}_{b3}(t)}\left[\frac{1}{2}\lVert\nabla\Phi_2\rVert^2\boldsymbol{n}_3-\frac{\partial\Phi_2}{\partial\,n_3}\nabla\Phi_2\right]\text{d}A\text{d}t.
\end{dcases}
\label{eq:ISigma}
\end{equation}
Please note that when the cavitation bubble contacts with the droplet, the boundaries $\upSigma_{b2}$ and $\upSigma_{b3}$ vary with time. Then the total Kelvin impulse reads $\boldsymbol{I}_{S}=\boldsymbol{I}_{b1}+\boldsymbol{I}_{b2}+\boldsymbol{I}_{b3}$.

Referring to \citet{SupponenObreschkow-152}, the Kelvin impulse is nondimensionalised as
\begin{equation}
    \boldsymbol{\zeta}=\frac{\boldsymbol{I}_{S}}{4.789\,R_{b,\max}^3\sqrt{\upDelta p\,\rho_{w}}},
    \label{eq:zeta_t}
\end{equation}
where $\boldsymbol{\zeta}$ is also called the anisotropy parameter. In the following sections, we mainly use the anisotropy parameter $\boldsymbol{\zeta}$ to describe the bubble centroid migration.

\subsection{Bubble motion during growth and collapse}

\begin{figure}
\centerline{\includegraphics[width=70 mm]{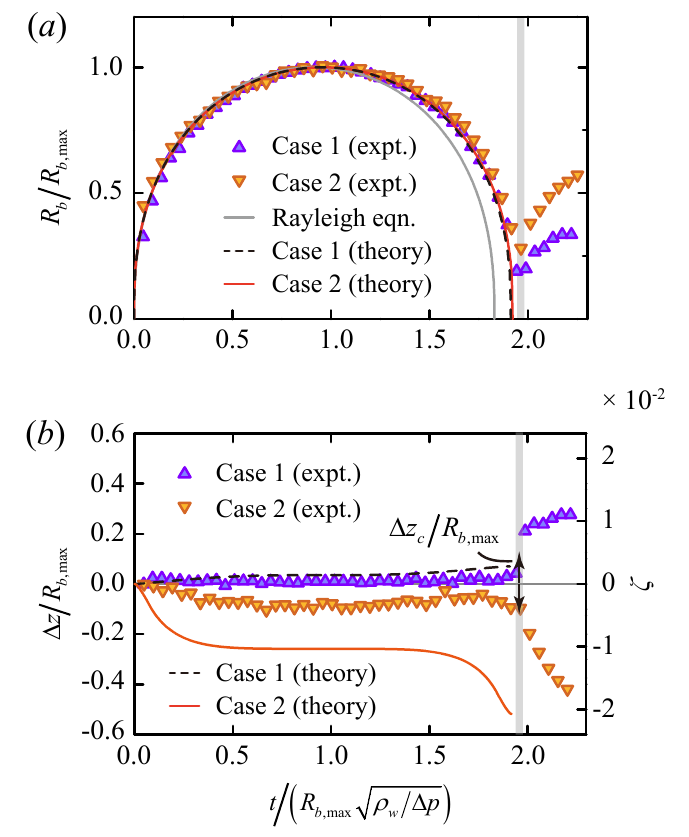}}
\caption{\label{fig:Impulse} Migration of the bubble centroid near the pendant kerosene droplet with density ratio $\rho_{o}/\rho_{w}=0.80$. (\textit{a}) Evolution of the dimensionless equivalent radius of the cavitation bubble $R_{b}/R_{b,\max}$ with dimensionless time $t/(R_{b,\max}\sqrt{\rho_{w}/\Delta p})$. (\textit{b}) Evolution of the dimensionless displacement of the bubble centroid in $z$ direction $\upDelta z/R_{b,\max}$ (ordinate on the left-hand side) and evolution of the anisotropy parameter $\zeta$ (ordinate on the right-hand side). Case 1 (expt.): $L/R_{d,0}=1.3\pm0.1$, $R_{b,\max}/R_{d,0}=0.31\pm0.03$. Case 2 (expt.): $L/R_{d,0}=1.14\pm0.03$, $R_{b,\max}/R_{d,0}=0.30\pm0.01$. Case 1 (theory): $L/R_{d,0}=1.40$, $R_{b,\max}/R_{d,0}=0.31$. Case 2 (theory): $L/R_{d,0}=1.14$, $R_{b,\max}/R_{d,0}=0.30$. In (\textit{b}), the data points are corresponding to the ordinate on the left-hand side, while the theoretical lines are corresponding to the ordinate on the right-hand side. The grey stripes in (\textit{a}) and (\textit{b}) denote the instants of the end of bubble collapses, when the dimensionless displacement of the cavitation bubble is defined as $\upDelta z_{c}/R_{b,\max}$ as denoted by the black arrows in (\textit{b}).}
\end{figure}

To verify the theoretical model for bubble dynamics, we compare the calculated anisotropy parameter with our experimental results. Figure~\ref{fig:Impulse}(\textit{a}) shows the evolution of the dimensionless bubble radius $R_{b}/R_{b,\max}$ with dimensionless time $t/(R_{b,\max}\sqrt{\rho_{w}/\Delta p})$ for cavitation bubbles generated near kerosene droplets. The bubble dynamics are assumed to follow the modified Rayleigh equation near a rigid boundary~\citep{BestBlake-212}, as shown below,
\begin{equation}
  R_b\,\ddot{R}_b+\frac{3}{2}\dot{R}_b^2+\frac{R_b}{2\,L}\left(R_b\,\ddot{R}_b+2\dot{R}_b^2\right)=-\frac{\upDelta p}{\rho_{w}},
  \label{eq:modRayleigh}
\end{equation}
with $\dot{R}_b$ and $\ddot{R}_b$ being the velocity and the acceleration of the bubble interface, respectively. 

Although equation (\ref{eq:modRayleigh}) does not include influences from the pendant droplets, the theoretical predictions show good agreement with the experimental results, see figure~\ref{fig:Impulse}(\textit{a}). The consistency between solutions to equation (\ref{eq:modRayleigh}) and experimental results indicates that the pendant droplets are not necessarily included in the bubble dynamics equations, probably because of the leading contribution of the infinitely large rigid boundary compared with the pendant droplet. Meanwhile, the solution to the Rayleigh equation~\citep{Rayleigh-301} is displayed in the grey line, which is theoretically valid for spherical bubble dynamics in an infinite liquid and underestimates the lifetimes of the bubbles in our cases. Here we should also note that equation (\ref{eq:modRayleigh}) is valid for relatively large $R_{b,\max}/R_{d,0}$ and large $\rho_o/\rho_w$ for the cases shown in our experiments.

The bubble centroid migration can be quantified with the dimensionless displacement of the bubble centre during growth and collapse, i.e., $\Delta z/R_{b,\max}$, as shown with respect to the ordinate on the left-hand side (l.h.s.) in figure~\ref{fig:Impulse}(\textit{b}). The two experimental cases displayed show different directions of bubble migration at the end of bubble collapse. In experimental case 1 (upper triangular markers), with $L/R_{d,0}=1.3\pm0.1$ and $R_{b,\max}/R_{d,0}=0.31\pm0.03$, the bubble centre suddenly migrates towards the solid boundary and the oil droplet at the end of collapse (grey stripe), which is consistent with previous observations \citep{SupponenObreschkow-152}. In experimental case 2 (lower triangular markers), with $L/R_{d,0}=1.14\pm0.03$ and $R_{b,\max}/R_{d,0}=0.30\pm0.01$, the motion of the bubble centre is transient with time, i.e., the bubble migrates away from the droplet during the growth, approaches the droplet during the collapse, and suddenly moves away from the droplet at the end of bubble collapse. The dimensionless displacements of the bubble at the end of the collapse are marked with the black arrows in figure~\ref{fig:Impulse}(\textit{b}), as defined by $\Delta z_{c}/R_{b,\max}$. 

The evolution of the anisotropy parameter $\zeta$ is shown corresponding to the ordinate on the right-hand side (r.h.s.) in figure~\ref{fig:Impulse}(\textit{b}), with two theoretical cases displayed. In theoretical case 1 (black dashed line), with $L/R_{d,0}=1.40$ and $R_{b,\max}/R_{d,0}=0.31$, the anisotropy parameter is always positive and increases with time. In theoretical case 2 (orange solid line), with $L/R_{d,0}=1.14$ and $R_{b,\max}/R_{d,0}=0.30$, the anisotropy parameter is always negative and shows similar trends to the evolution of the bubble centre in experimental case 2. The complex evolution may be related to the contact between the bubble and the droplet. In case 1 (both the experimental and the theoretical results), the bubble does not contact the droplet during its first growth and collapse, and the attractive force from the rigid boundary plays a leading role, thus causing a relatively smooth evolution of the displacement and the anisotropy parameter. By contrast, in case 2 (both the experimental and the theoretical results), the bubble contacts with the oil droplet during its growth, leading to an increasing repulsive force from the oil droplet on the bubble due to the component Kelvin impulse $\boldsymbol{I}_{b2}$ from the droplet-water interface $\upSigma_{b2}$ and thus the bubble motion away from the oil droplet. During the bubble collapse, after the bubble detaches from the droplet interface, the attractive force from the rigid boundary causes the upward motion of the bubble again. At the end of the bubble collapse, the impulse of the repulsive force dominates over the impulse of the attractive force, thus leading to a negative displacement of the bubble.

\subsection{Bubble centroid displacement at the end of collapse}
\label{sec:bubble_displacement}

\begin{figure}
\centerline{\includegraphics[width=130 mm]{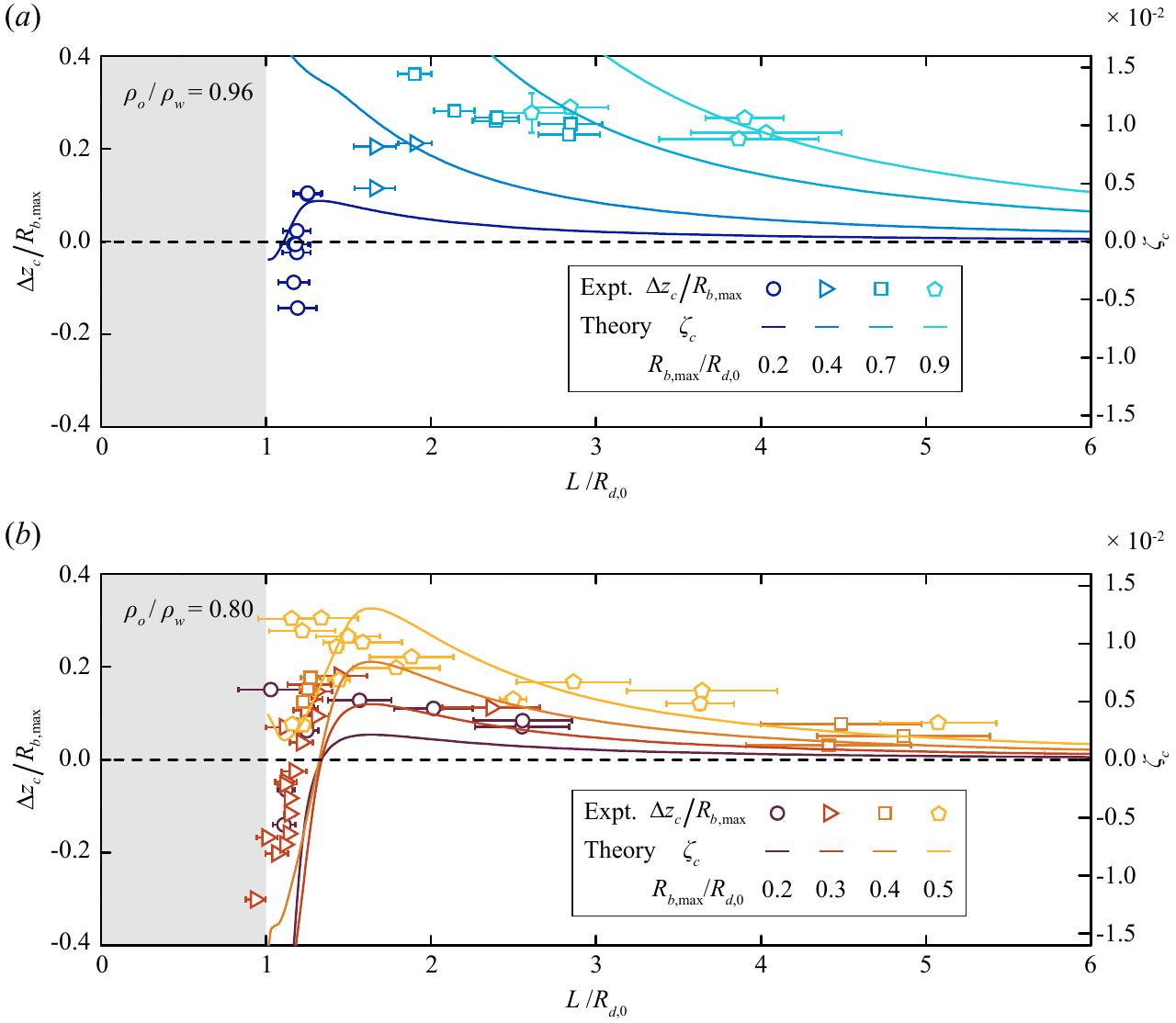}}
\caption{\label{fig:zeta_Lw} Dimensionless displacement of the bubble centroid at the end of collapse $\Delta z_{c}/R_{b,\max}$ as a function of $L/R_{d,0}$ (ordinate on the left-hand side, data points), and the anisotropy parameter at bubble collapse $\zeta_{c}$ as a function of $L/R_{d,0}$ (ordinate on the right-hand side, solid lines) for interactions of cavitation bubbles with (\textit{a}) silicone oil droplets ($\rho_{o}/\rho_{w}=0.96$) and (\textit{b}) kerosene droplets ($\rho_{o}/\rho_{w}=0.80$) with different size ratios $R_{b,\max}/R_{d,0}$. The shaded areas denote the interior of the droplet.}
\end{figure}

The variations of $\Delta z_{c}/R_{b,\max}$ with $L/R_{d,0}$ and $R_{b,\max}/R_{d,0}$ are shown near silicone oil droplets in figure~\ref{fig:zeta_Lw}(\textit{a}) and near kerosene droplets in figure~\ref{fig:zeta_Lw}(\textit{b}). The ordinate on the l.h.s. denotes $\Delta z_{c}/R_{b,\max}$, while the ordinate on the r.h.s. denotes $\zeta_{c}$, which is defined as the anisotropy parameter at the end of the bubble collapse.

For silicone oil droplets, as shown in figure~\ref{fig:zeta_Lw}(\textit{a}), four different size ratios $R_{b,\max}/R_{d,0}$ are selected, i.e., $R_{b,\max}/R_{d,0}=0.2, 0.4, 0.7,$ and 0.9, with the dimensionless distance $L/R_{d,0}=1.1$ -- 4.0. The experimental data indicate that with $R_{b,\max}/R_{d,0}=0.2$ the critical dimensionless distance $L/R_{d,0}$ is around 1.1 for the conversion of the bubble migration direction, which is consistent with the theoretical prediction at $\zeta_{c}=0$. Besides, our theoretical model predicts that with $R_{b,\max}/R_{d,0}=0.4$, the bubble migrates towards the rigid boundary at the end of the collapse for all $L/R_{d,0}>1$. With large $R_{b,\max}/R_{d,0}$, the invariability of the bubble migration direction is attributed to the component contribution of $\boldsymbol{I}_{b3}$ from the bubble-droplet interface $\upSigma_{b3}$, which is discussed in more detail in appendix~\ref{sec:app_theory}. 

For kerosene droplets, as shown in figure~\ref{fig:zeta_Lw}(\textit{b}), four different size ratios $R_{b,\max}/R_{d,0}$ are selected, i.e., $R_{b,\max}/R_{d,0}=0.2, 0.3, 0.4,$ and 0.5, with the dimensionless distance $L/R_{d,0}=0.9$ -- 5.1. The experimental data show that with $R_{b,\max}/R_{d,0}=0.2$ -- 0.4 the critical $L/R_{d,0}$ for the conversion of the bubble motion direction is around 1.2 -- 1.3. Within the same range of $R_{b,\max}/R_{d,0}$, the theoretical curves of $\zeta_{c}$ predict the same critical $L/R_{d,0}\approx\,1.33$. With $R_{b,\max}/R_{d,0}=0.5$, the bubble migrates towards the rigid boundary at the end of the collapse for all $L/R_{d,0}>1$. The reason for the critical $L/R_{d,0}$ in the theory slightly larger than in the experiment could be related to the sphericity of the kerosene droplet, see the discussions in appendix~\ref{sec:app_expt}.

In short, the discussions above have verified that the anisotropy parameter calculated from our theory can be applied to quantifying the bubble centroid migration at the end of the collapse, which is used to explain the different interactions of the bubble and the droplet in \S\ref{sec:discussion}. 

\section{Regimes of bubble-droplet interactions}
\label{sec:discussion}

In \S\ref{sec:exptresults}, we have already shown the overview of the four regimes of bubble-droplet interactions, namely, the oil droplet rupture, the water droplet entrapment, the oil droplet large deformation, and the oil droplet mild deformation. In this section, we show more details on the flow induced by the cavitation bubble after the collapse and the responses of the droplet to the flow. Finally, we propose a phase diagram for the regimes by analysing the droplet dynamics. 

\subsection{Regime 1: Oil droplet rupture}
\label{sec:discussion_rupture}

\begin{figure}
	\centerline{\includegraphics[width=130 mm]{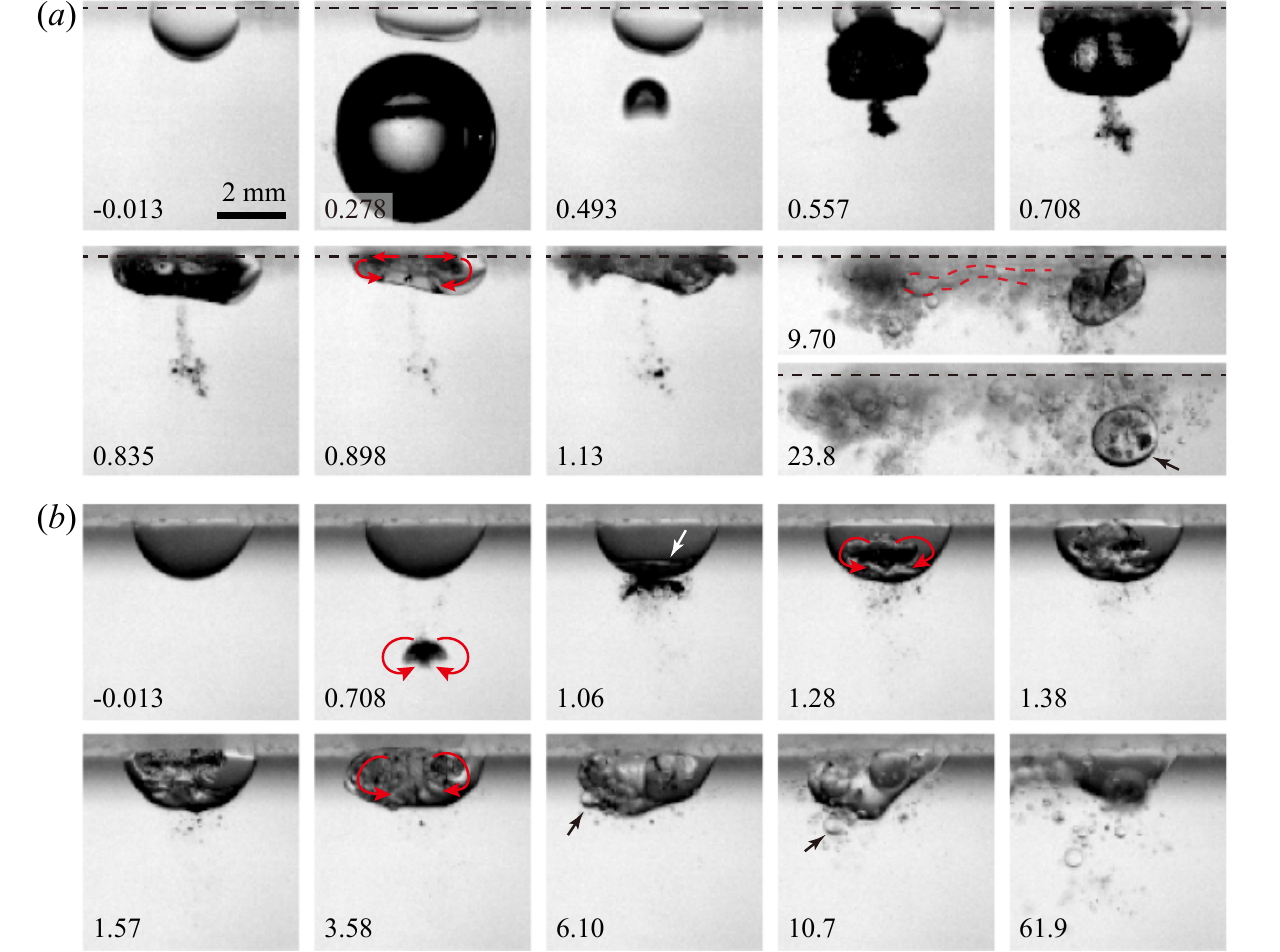}}
	\caption{\label{fig:Response_rupture} Details of oil droplet rupture. Silicone oil droplets are ruptured due to (\textit{a}) bubble jet impact at $L/R_{d,0}=2.7\pm0.3$, $R_{b,\max}/R_{d,0}=1.6\pm0.2$, and (\textit{b}) bubble vortex ring impact at $L/R_{d,0}=3.3\pm0.1$, $R_{b,\max}/R_{d,0}=1.30\pm0.05$. Photographs in (\textit{a}) and (\textit{b}) share the same scale bar length of 2 mm. The times are in the units of ms with 0 ms for the laser-plasma generation. The red arrows indicate the flow directions, the white arrow at 1.06 ms in (\textit{b}) denotes the interface, and the black arrows denote the pinched-off oil droplets. The movies are integrated and provided online as supplementary movie 3 at https://doi.org/xxxx.}
\end{figure}

Two typical ways are observed in our experiments to realise the oil droplet rupture, namely, by bubble jet impact and by bubble vortex ring impact, see figure~\ref{fig:Response_rupture}. 

In figure~\ref{fig:Response_rupture}(\textit{a}), near a silicone oil droplet with $L/R_{d,0}=2.7\pm0.3$ and $R_{b,\max}/R_{d,0}=1.6\pm0.2$, the cavitation bubble generates a pronounced jet after collapse (0.493 ms) which directly impacts the droplet during the first rebound of the bubble (0.557 ms). The rebounding bubble enters and penetrates the oil droplet before it impacts the rigid boundary (0.835 ms). Then the bubble evolves into a bubble vortex ring (as denoted by the arrows at 0.898 ms) and induces strong shear flows along the rigid boundary~\citep{ZengAn-454}. The expansion and circulation of the bubble vortex ring exert strong tensile and shear stresses on the oil droplet, which can be visualised by the oil ligament as outlined in the red dashed lines at 9.70 ms. In this way, the oil droplet is ruptured into multiple daughter droplets, with radii $\lesssim\,120\,\upmu$m, which is close to emulsification. The daughter droplet denoted by the arrow at 23.8 ms contains entrained water droplets and gaseous bubble remnants, indicating the formation of W/O/W (water in oil in water) structures. Moreover, in this case, the oil droplet is detached from the rigid boundary, thus realising the removal of the pendant oil droplet by the jetting of the cavitation bubble.

The rupture of the silicone oil droplet by bubble vortex ring impact is shown in figure~\ref{fig:Response_rupture}(\textit{b}), with $L/R_{d,0}=3.3\pm0.1$ and $R_{b,\max}/R_{d,0}=1.30\pm0.05$. The bubble vortex ring is generated after the bubble collapse as denoted by the arrows at 0.708 ms and then it translates upwards due to initial impulse and buoyancy. Before the collision of the bubble vortex ring and the oil droplet, a dimple is already seen through the droplet, indicating that a water column is driven by the motion of the vortex ring, as denoted by the arrow at 1.06 ms. The vortex ring enters the droplet and circulates (1.28 ms) before it impacts the rigid boundary (1.57 ms) and expands radially, see the arrows at 3.58 ms. The stretching and shearing are similar to the case in figure~\ref{fig:Response_rupture}(\textit{a}), although the strength is much weaker due to the smaller impulse of the jetting bubble. The circulation of the vortex ring causes the formation and rupture of oil ligaments, and multiple daughter oil droplets with radii $\lesssim\,600\,\upmu$m are pinched off (e.g., see arrows at 6.10 ms and 10.7 ms). The pendant oil droplet is not totally removed, but it loses about one-third of the weight in the rupture process.

In experiments, the contact line of the droplet is observed to be pinned at the PMMA
substrate during the impact of the water jet; see figure~\ref{fig:Response_rupture}. When the bubble vortex ring collides with the substrate and expands to the rim of the droplet, it drives the slippage of the droplet contact line; see frames at 1.13 ms in figure~\ref{fig:Response_rupture}(\textit{a}) and at 3.58 ms in figure~\ref{fig:Response_rupture}(\textit{b}). Finally, the droplet is either detached from the substrate (figure~\ref{fig:Response_rupture}(\textit{a})) or contracts to a smaller one but with a similar contact angle to the original one (figure~\ref{fig:Response_rupture}(\textit{b})), depending on the strength of the bubble jet. 

\subsection{Regime 2: Water droplet entrapment}

Water droplet entrapment occurs when an upward jetting flow of water enters the oil droplet and pinches off, as already shown in figure~\ref{fig:ExpResults_SiO}(\textit{b}) and figure~\ref{fig:ExpResults_WK}. Here we summarise two main ways to realise the water droplet entrapment, as follows.

\begin{figure}
	\centerline{\includegraphics[width=130 mm]{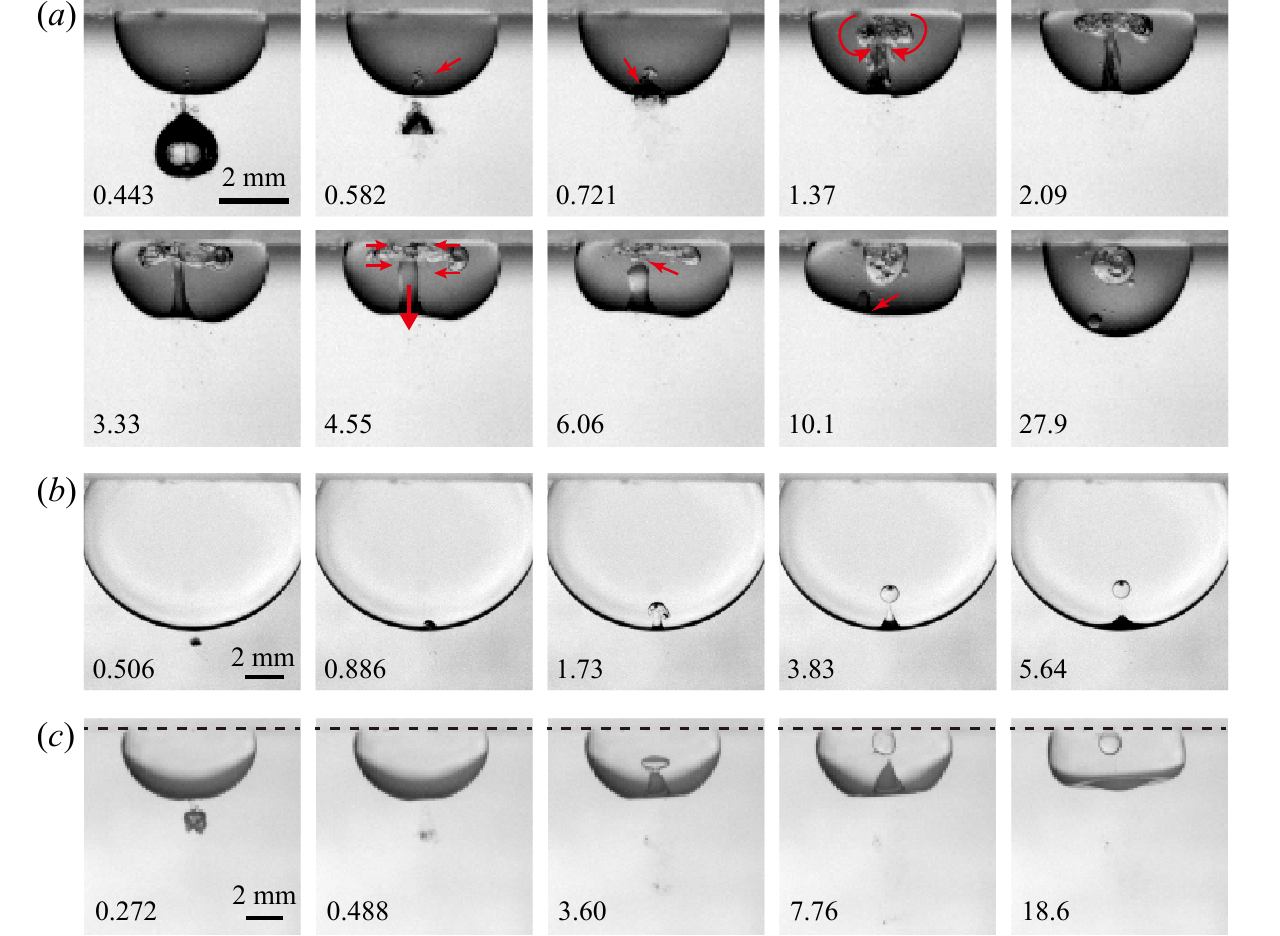}}
	\caption{\label{fig:Response_entrap} Details of water droplet entrapment in the oil droplet after bubble collapse. With a silicone oil droplet, cases are displayed at (\textit{a}) $L/R_{d,0}=2.1\pm0.1$, $R_{b,\max}/R_{d,0}=0.69\pm0.04$, 
    and (\textit{b}) $L/R_{d,0}=1.4\pm0.1$, $R_{b,\max}/R_{d,0}=0.27\pm0.03$. (\textit{c}) With a kerosene droplet, one case is displayed at $L/R_{d,0}=1.14\pm0.03$, $R_{b,\max}/R_{d,0}=0.30\pm0.01$. 
    The times are in the units of ms with 0 ms for the laser-plasma generation. The arrows indicate the flows at 1.37 ms and 4.55 ms in (\textit{a}) and the interfaces in the other frames. The movies are integrated and provided online as supplementary movie 4 at https://doi.org/xxxx.}
\end{figure}

The first way is characterised by the pinch-off of an upward water column during the contract in the oil droplet, as shown in figure~\ref{fig:Response_entrap}(\textit{a}). After the bubble collapses and generates a micro-jet during rebound (0.443 ms), the bubble jet penetrates the silicone oil droplet (arrow at 0.582 ms) and makes the bubble evolve into a bubble vortex ring, which drives an upward water column as a dimple at the bottom of the oil droplet (arrow at 0.721 ms). Then the water column moves upwards while circulating (arrows at 1.37 ms) before it reaches the rigid boundary (2.09 ms) and spreads radially (3.33 ms). As the kinetic energy dissipates, gravity dominates again and drives the contract of the water column (arrows at 4.55 ms) which pinches off and leaves a large water droplet at the top (arrow at 6.06 ms). Interestingly, the contract of the water column drives a second pinch-off (arrow at 10.1 ms), thus entrapping multiple water droplets inside the oil droplet. 

The second way is characterised by the pinch-off of an upward water column during ascending, which can be realised by the bubble motion either upwards (figure~\ref{fig:Response_entrap}(\textit{b})) or downwards (figure~\ref{fig:Response_entrap}(\textit{c})). The pinch-off of the water column induced by a downward bubble jet has already been reported by \citet{HanZhang-116}, where a cavitation bubble is initiated near a flat oil-water interface. Here the pinch-off of the water column may be explained by the oscillation in the surface energy of a cylindrical column which tends to be magnified and generates daughter droplets, known as Rayleigh-Plateau instability~\citep{Chandrasekhar-333}. The criteria of the instability will be applied to the analysis of the pinch-off in \S\ref{sec:Regime_phase}.

\subsection{Regime 3: Oil droplet large deformation}

\begin{figure}
	\centerline{\includegraphics[width=130 mm]{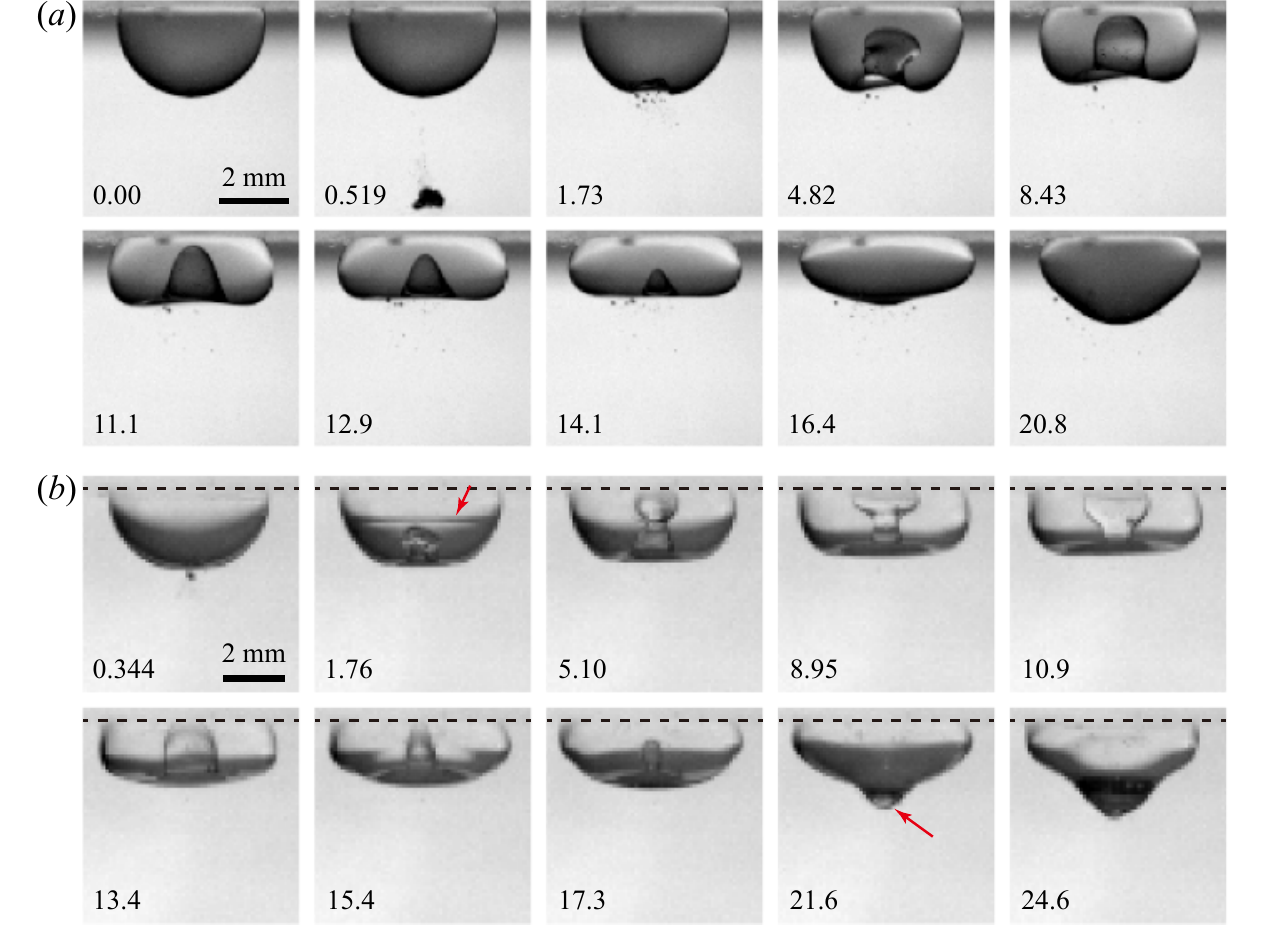}}
	\caption{\label{fig:Response_deformation} Details of oil droplet large deformation after bubble collapse. With a silicone oil droplet, one case is displayed at (\textit{a}) $L/R_{d,0}=2.8\pm0.2$, $R_{b,\max}/R_{d,0}=0.70\pm0.05$. With a kerosene droplet, one case is displayed at (\textit{b}) $L/R_{d,0}=1.23\pm0.04$, $R_{b,\max}/R_{d,0}=0.46\pm0.02$. 
    The times are in the units of ms with 0 ms for the laser-plasma generation. The arrows indicate the capillary wave at 1.76 ms in (\textit{b}) and the interface at 21.6 ms in (\textit{b}). The movies are integrated and provided online as supplementary movie 5 at https://doi.org/xxxx.}
\end{figure}

The oil droplet's large deformation refers to the visibility of an upward water column at the bottom of the oil droplet which does not pinch off. Two main ways are observed for silicone oil and kerosene droplets, as summarised in figure~\ref{fig:Response_deformation}. 

The first way to realise large deformation of the oil droplet has already been shown in figure~\ref{fig:ExpResults_SiO}(\textit{c}). The details are shown here in figure~\ref{fig:Response_deformation}(\textit{a}). The water column ascends in the silicone oil droplet and reaches a maximum height \textit{without touching} the rigid boundary (8.43 ms). The behaviours of the water column during ascending are similar to the cases through a flat oil-water interface~\citep{HanZhang-116}. Different from previous studies, during descending, the water column evolves into a conical shape while keeping the bottom of the oil droplet flat (11.1 ms to 14.1 ms). Then the droplet oscillates until it recovers to its original state. 

The second way to realise large deformation of the oil droplet is illustrated in figure~\ref{fig:Response_deformation}(\textit{b}), where the water column ascends inside the kerosene droplet, \textit{touches} the rigid boundary (5.10 ms) and spreads radially (8.95 ms). Then the oil droplet oscillates with its bottom lifted to enhance the contraction of the water column (10.9 ms). The shape of the water column evolves from a thick cylinder (13.4 ms) to a cone (15.4 ms), and finally to a thin cylinder (17.3 ms). Especially, the thin cylinder pinches off a small water droplet (arrow at 21.6 ms). With a very thin oil gap, the small water droplet merges with the bulk water only within 3 ms, followed by oil droplet oscillations until recovery.    

\subsection{Regime 4: Oil droplet mild deformation}
\label{sec:discussion_mild}

\begin{figure}
	\centerline{\includegraphics[width=130 mm]{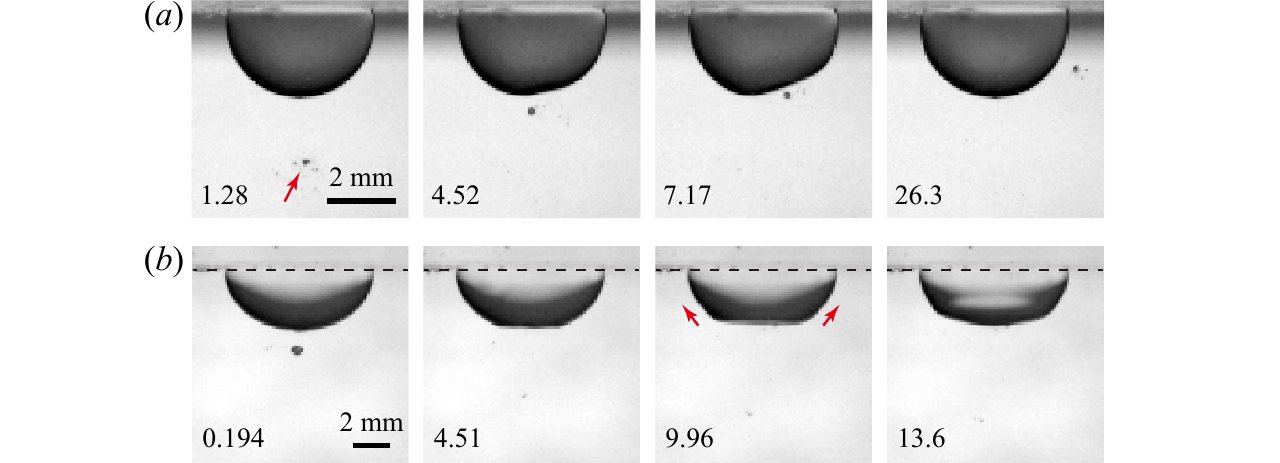}}
	\caption{\label{fig:Response_mildresponse} Details of oil droplet mild deformation after bubble collapse. With a silicone oil droplet, one case is displayed at (\textit{a}) $L/R_{d,0}=2.9\pm0.2$, $R_{b,\max}/R_{d,0}=0.57\pm0.04$. With a kerosene droplet, one case is displayed at (\textit{b}) $L/R_{d,0}=1.11\pm0.07$, $R_{b,\max}/R_{d,0}=0.24\pm0.02$. The times are in the units of ms with 0 ms for the laser-plasma generation. The arrows indicate the bubble remnants at 1.28 ms in (\textit{a}) and the flow direction at 9.96 ms in (\textit{b}). The movies are integrated and provided online as supplementary movie 6 at https://doi.org/xxxx.}
\end{figure}

The oil droplet's mild deformation refers to the phenomenon that no water column is visible inside the oil droplet. Two examples are shown in figure~\ref{fig:Response_mildresponse} for a silicone oil droplet and a kerosene droplet. 

In figure~\ref{fig:Response_mildresponse}(\textit{a}), the silicone oil droplet induces an upward motion of the collapsing cavitation bubble with $L/R_{d,0}=2.9\pm0.2$ and $R_{b,\max}/R_{d,0}=0.57\pm0.04$. The flow can be visualised by the bubble remnants as marked by the arrow at 1.28 ms. The upward flow is not strong enough to generate a water column, but only slightly deforms the droplet (7.17 ms). Therefore, the bubble remnants move around the droplet towards the rigid boundary (26.3 ms). 

In figure~\ref{fig:Response_mildresponse}(\textit{b}), the kerosene droplet induces a downward motion of the collapsing cavitation bubble with $L/R_{d,0}=1.11\pm0.07$ and $R_{b,\max}/R_{d,0}=0.24\pm0.02$. The focused water flow between the bubble remnants and the oil droplet collides with the oil droplet and generates surface waves propagating along the oil-water interface (arrows at 9.96 ms). The surface waves have also been observed in previous regimes (arrow at 1.76 ms in figure~\ref{fig:Response_deformation}(\textit{b})). 

\subsection{Phase diagram}
\label{sec:Regime_phase}

Section~\ref{sec: modelling} shows that the anisotropy parameter at bubble collapses $\zeta_{c}$ can quantitatively reflect the bubble centroid migration, including the direction and the strength. Although the physics is complex during the interaction of the water jet with the oil droplet, in this section, we tend to extract the main mechanisms that control the dynamics of the water jet and ignore the other minor effects to obtain a general picture of the phase diagram.

The detailed investigations of the different bubble-droplet interactions in \S\ref{sec:discussion_rupture} to \S\ref{sec:discussion_mild} suggest that (\textit{i}) the phenomena of oil droplet's deformation and water droplet's entrapment are classified by the pinch-off of the water column penetrating the oil droplet, and (\textit{ii}) the determination of water droplet's entrapment and oil droplet's rupture is probably related to the size (surface area) of the inward water column. 

First, we explain the critical condition for the oil droplet's deformation and the water droplet's entrapment. Similar to the theoretical model proposed by \citet{HanZhang-116}, an upward water column is assumed to ascend from the bottom of the oil droplet with an initial linear momentum $I_{w}$. The surface tension dominates over the gravitational effects, with a Bond number $\textit{Bo}_1=(\rho_w-\rho_o)\,g\,R_{b,\max}^2/\sigma\lesssim\,0.05\ll\,1$, with $\rho_w=1\times\,10^3\,\text{kg}/\text{m}^3$, $\rho_o\gtrsim\,8\times\,10^2\,\text{kg}/\text{m}^3$, $g=9.81\,\text{m}/\text{s}^2$, $R_{b,\max}\approx\,1\times\,10^{-3}\,$m, and $\sigma\approx\,4\times\,10^{-2}\,\text{N}/\text{m}$. To assess the effect of the viscosity, we estimate the Reynolds number $\textit{Re}=u_m\,R_{b,\max}/\nu_o$, with the characteristic velocity of the water column $u_m\gtrsim\,0.1\,$m/s, the length scale $R_{b,\max}\approx\,10^{-3}\,$m, and the kinematic viscosity $\nu_o\approx\,5\times\,10^{-5}\,\text{m}^2/\text{s}$ for silicone oil droplets and $\nu_o\approx\,2\times\,10^{-6}\,\text{m}^2/\text{s}$ for kerosene droplets. Therefore for silicone oil droplets, the Reynolds number is larger than 2, while for kerosene droplets the Reynolds number is larger than 50, indicating that the viscosity plays a limited role during the evolution of the water jet. From a view of energy balance, the viscous force leads to dissipation, and the surface force leads to an increase in surface energy. Therefore the maximum height $h_m$ of the water column should be determined by the balance of the surface energy of the water column $E_{s1}$ and the kinetic energy of the water jet $E_{k}$, with $E_{s1}\approx\,\sigma\,2\,\upi\,R_{b,\max}\,h_m$ and $E_{k}\approx\,\frac{1}{2}\,M_w\,u_w^2$. The mass of the water jet is estimated as $M_w\approx\,\rho_w\,\upi\,R_{b,\max}^2\,h_m$ and the velocity $u_w$ should be on a scale of $I_w/M_w$. The linear momentum $I_{w}$ is on the same magnitude of the Kelvin impulse at the bubble collapse $I_{S,c}$ which is transferred by the bubble jet or by the bubble vortex ring, i.e., $I_{w}\sim\,I_{S,c}$. Using (\ref{eq:zeta_t}) and introducing $\zeta_{c}$, we obtain
\begin{equation}
h_{m}\propto\,\sqrt{\frac{\rho_w\,u_0^2\,R_{b,\max}^3}{\sigma}\,\zeta_c^2},
\end{equation}
where $u_0$ is the velocity scale defined as $u_0=\sqrt{\Delta p/\rho_{w}}$. 

\begin{figure}
	\centerline{\includegraphics[width=90 mm]{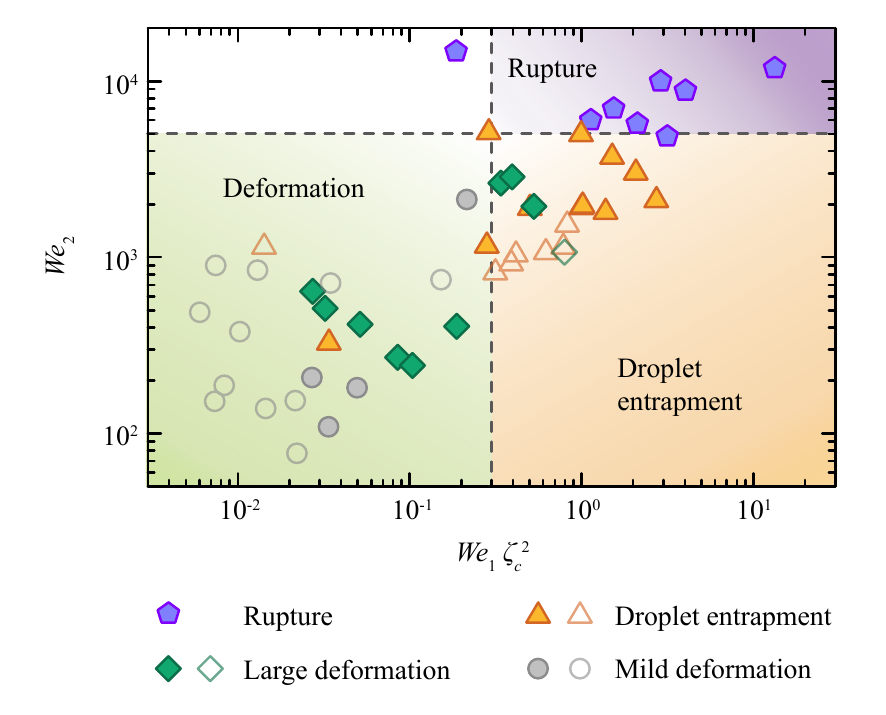}}
	\caption{\label{fig:phase_diagram} Phase diagram of cavitation bubble interactions with silicone oil droplets (filled markers) and kerosene droplets (empty markers) determined by the Weber number $\textit{We}_2=(\rho_w\,u_0^2\,R_{b,\max}^3)/(\sigma\,R_{d,0}^2)$ vs.~$\textit{We}_1\,\zeta_{c}^2$ with the Weber number $\textit{We}_1=(\rho_w\,u_0^2\,R_{b,\max})/\sigma$ and the velocity scale $u_0=\sqrt{\Delta p/\rho_w}$. The dashed lines guide the divisions for bubble interactions with silicone oil droplets, leading to three coloured regions for oil droplet responses, namely, rupture, (water) droplet entrapment, and deformation.}
\end{figure}

From the criteria of Rayleigh-Plateau instability~\citep{Charru-60}, the water column may pinch off when its length reaches the wavelength of the interfacial perturbation and becomes larger than the perimeter of the cross-section, leading to the comparison between the longitudinal length $h_{m}$ and the transversal length $R_{b,\max}$. Then we obtain $h_{m}/R_{b,\max}\propto\,\sqrt{\textit{We}_1\,\zeta_c^2}$, with the Weber number being $\textit{We}_1=\rho_w\,u_0^2\,R_{b,\max}/\sigma$. The pinch-off cases, i.e., the water droplet entrapment, require $\textit{We}_1\,\zeta_{c}^2$ larger than a constant, which is guided by the vertical dashed line for silicone oil droplets (filled markers) in figure~\ref{fig:phase_diagram}. For kerosene droplets (empty markers), the critical $\textit{We}_1\,\zeta_{c}^2$ slightly decreases probably because the viscous dissipation of the ascending water column is less important both in kerosene and in silicone oil with a low viscosity.

Finally, we clarify the critical condition for oil droplet rupture and water droplet entrapment. The total energy of a laser-induced cavitation bubble $E_0$ can be characterised by the potential energy of the bubble at its maximum size~\citep{TinguelyObreschkow-539}, as follows,
\begin{equation}
    E_0=\frac{4\,\upi}{3}\,\Delta p\,R_{b,\max}^3.
\end{equation}
After the pinch-off of a single daughter water droplet with radius $R_{d,p}$ inside the oil droplet, the increase of the surface energy $E_{s2}$ reads
\begin{equation}
    E_{s2}={4\,\upi}\sigma\,R_{d,p}^2,
\end{equation}
with $\sigma$ the surface tension coefficient between water and oil. Here the increase of the potential energy is ignored because the water droplet (e.g.~figure~\ref{fig:ExpResults_WK}(\textit{b}) and~\ref{fig:ExpResults_WK}(c)), at the quasi-steady state, can rest just above the bottom of the oil droplet, with a Bond number $\textit{Bo}_2=(\rho_w-\rho_o)\,g\,R_{d,p}^2/\sigma\lesssim\,0.05\ll\,1$, with $R_{d,p}\sim\,1\times\,10^{-3}\,$m. 

Since the total energy $E_0$ can dissipate as shock waves~\citep{TinguelyObreschkow-539} and work done by viscous forces, the surface energy $E_{s2}$ would be part of $E_0$, and reasonably increases with increasing $E_0$, i.e., $E_{s2}\propto E_0$. The oil droplet would rupture when the size of the daughter water droplet is comparable to the oil droplet, thus leading to $R_{d,p}/R_{d,0}$, which can be simplified as the Weber number being {$\textit{We}_2=(\rho_w\,u_0^2\,R_{b,\max}^3)/(\sigma\,R_{d,0}^2)$.} A larger Weber number $\textit{We}_2$ corresponds to daughter droplets with larger sizes, as shown by the horizontal dashed line dividing the regimes of oil droplet rupture and water droplet entrapment for silicone oil droplets in figure~\ref{fig:phase_diagram}.

The phase diagram in figure~\ref{fig:phase_diagram} clearly shows that the different responses of the oil droplets can be classified by two dominating non-dimensional parameters, $\textit{We}_1\,\zeta_c^2$ and $\textit{We}_2$. For silicone oil droplets, the regime of droplet deformation (including large and mild deformation) can be observed at $\textit{We}_1\,\zeta_c^2\,\lesssim\,0.35$ and $\textit{We}_2\,\lesssim\,5\times\,10^3$. By adjusting $L/R_{d,0}$ and $R_{b,\max}/R_{d,0}$, the anisotropy parameter at bubble collapse $\zeta_c$ can be adjusted, according to figure~\ref{fig:zeta_Lw}. Thus by controlling $\textit{We}_2\,\lesssim\,5\times\,10^3$ while increasing $\textit{We}_1\,\zeta_c^2$ to over 0.35, the regime transitions from oil droplet deformation to water droplet entrapment. Further increasing $\textit{We}_2$ to over $5\times\,10^3$, the regime of oil droplet rupture can be observed. In summary, the proposed phase diagram provides a simple way to identify the parameter space for desired regimes of droplet responses, as required in ultrasonic cleaning or emulsification.
 
\section{Conclusions}

In conclusion, we study experimentally and theoretically the interactions of a collapsing laser-induced cavitation bubble with a hemispherical droplet attached to a rigid boundary. In experiments, an approximately hemispherical droplet of silicone oil or kerosene is attached to the bottom surface of a fixed PMMA plate immersed in water. A laser-induced cavitation bubble is generated below the pendant oil droplet and the bubble-droplet interactions are recorded with high-speed imaging. By controlling the dimensionless distance from the centre of the cavitation bubble to the rigid boundary $L/R_{d,0}$ and the radius ratio $R_{b,\max}/R_{d,0}$, we observe four typical interactions between cavitation bubbles and pendant oil droplets, namely, the oil droplet rupture, the water droplet entrapment, the oil droplet large deformation, and the oil droplet mild deformation. In the first two regimes of interactions, emulsification of the oil and water droplets is observed.

The bubble dynamics are decisive to the understanding of the bubble-droplet interactions. Since previous models have not considered the influences of a curved liquid-liquid interface on bubble dynamics, we propose a new model with the method of images to quantitatively describe the bubble centroid migration at the end of bubble collapse. By calculating the anisotropy parameter, our model successfully predicts the critical dimensionless bubble-wall distances for the conversion of bubble migration direction with small bubble-droplet size ratios. We also prove theoretically that for large bubble-droplet size ratios, the bubble only migrates towards the rigid boundary at collapse, which agrees well with experiments. 

Finally, we investigate in detail the different ways to realise each regime of bubble-droplet interactions. We propose the critical conditions for the divisions of oil droplet deformation, water droplet entrapment, and oil droplet rupture, by illustrating the different regimes in a phase diagram with the combination of the Weber number and the anisotropy parameter.

Future work may focus on two aspects. First, the contact angles of the pendant droplet can be adjusted to find the influences of droplet shapes on the bubble-droplet interactions. Second, the physicochemical properties of the droplet and the bulk liquid (viscosity, surface tension, solubility, etc.) may also be varied to broaden the conclusions of the current research. Our findings may inspire the removal of sessile or pendant oil droplets, emulsification, cell rupture, and drug delivery by needle-free jet injections.

\backsection[Supplementary movies]{Supplementary movies are available at https://doi.org/xxxx.}
 
\backsection[Acknowledgements]{We gratefully acknowledge Y.~Guo for the technical support and Prof.~C.D.~Ohl for helpful discussions.}

\backsection[Funding]{This work was supported by the National Natural Science Foundation of China (grant nos.~52076120,~52079066, and~11988102); the State Key Laboratory of Hydroscience and Engineering (grant nos.~2019-KY-04, sklhse-2020-E-03 and, sklhse-2020-E-05); the Creative Seed Fund of Shanxi Research Institute for Clean Energy, Tsinghua University; and the China Postdoctoral Science Foundation (H.~Han, grant no.~2022M711766).}

\backsection[Declaration of interests]{The authors report no conflict of interest.}

\backsection[Author ORCIDs]{
\\Z. Ren, https://orcid.org/0000-0002-4682-9274; \\H. Han, https://orcid.org/0000-0003-0225-5624;\\H. Zeng, https://orcid.org/0000-0003-3190-5892;\\C. Sun, https://orcid.org/0000-0002-0930-6343;\\Y. Tagawa, https://orcid.org/0000-0002-0049-1984;\\ Z. Zuo, https://orcid.org/0000-0002-7407-0904; \\ S. Liu, https://orcid.org/0000-0003-2525-2303.}

\appendix

\section{Sphericity of the oil droplets and eccentricity of the bubble-droplet pairs in experiments}\label{sec:app_expt}

\begin{figure}
\centerline{\includegraphics[width=130 mm]{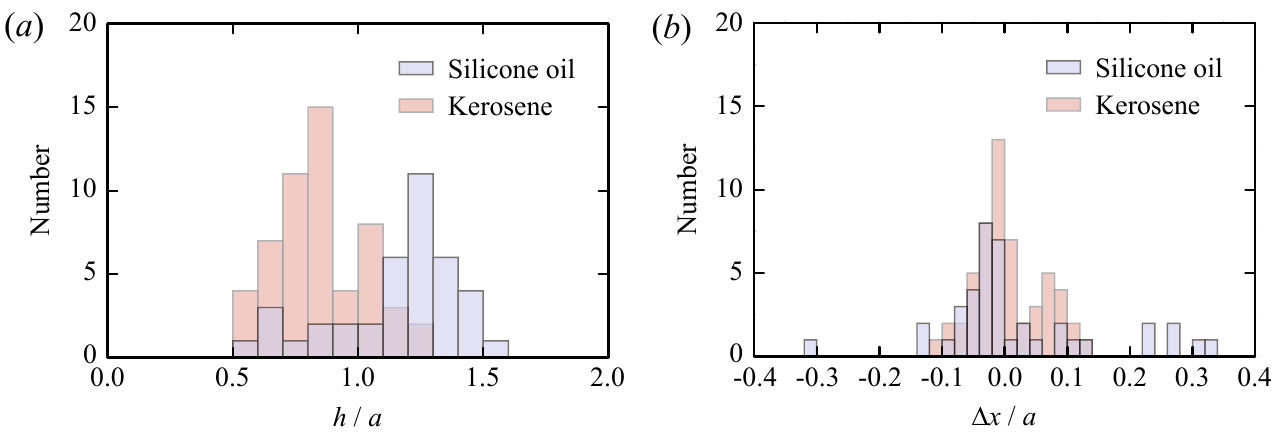}}
\caption{\label{fig:Sphericity_expt} Sphericity and eccentricity of the oil droplets. (\textit{a}) Distribution of the sphericity of oil droplets. (\textit{b}) Distribution of eccentricity of bubble-droplet pairs illustrated with the ratio $\Delta x/a$, where $\Delta x$ is the distance from the centre of the bubble to the symmetric axis of the oil droplet.}
\end{figure}

Figure~\ref{fig:Sphericity_expt}(\textit{a}) shows the distribution of the sphericity of the oil droplets used in experiments, as illustrated by the ratio of the droplet thickness $h$ and the droplet contact radius $a$. For silicone oil droplets, more cases lie in $h/a>1$, indicating that the droplet is longer in the vertical direction than in the horizontal direction. By contrast, for kerosene droplets, more cases lie in $h/a<1$, for which the effective radius $R_{d,0}$ is larger than the droplet thickness $h$, leading to the possibility of the bubble-wall distance $L\leqslant\,R_{d,0}$ in experiments. However, in theory, $L$ must be larger than $R_{d,0}$. This may explain the bias between experimental and theoretical results shown in figure~\ref{fig:zeta_Lw}(\textit{b}).

Figure~\ref{fig:Sphericity_expt}(\textit{b}) shows the distribution of the eccentricity of the oil droplets used in experiments, which is quantified with $\Delta x/a$, with $\Delta x$ being the distance from the centre of the bubble to the symmetric axis of the oil droplet and $a$ being the contact radius. For both types of droplets, $|\Delta x/a|$ mainly ranges within 10\%. Therefore we reckon this condition as the location of the cavitation bubble right below the pendant oil droplet.

\section{Verification of the theoretical model based on the method of images}\label{sec:app_image}

\begin{figure}
    \centering
    \includegraphics[width=130 mm]{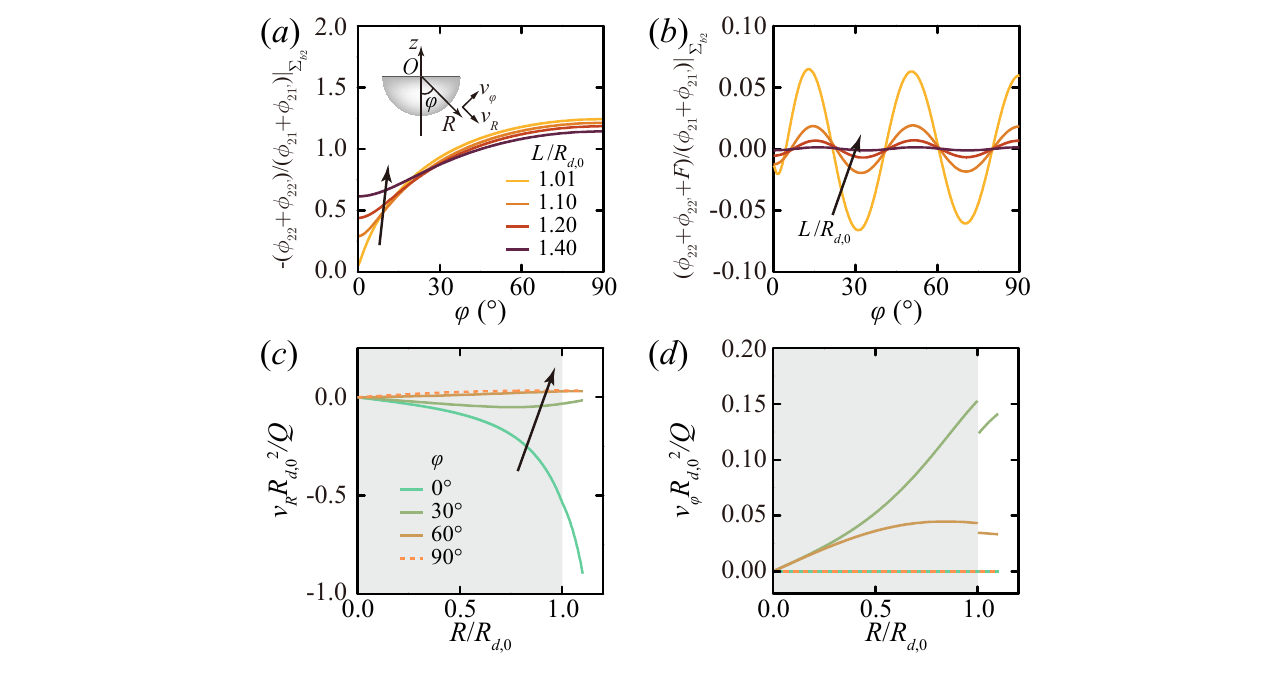}
    \caption{Verification of the theoretical model. (\textit{a}) Ratio between $-\left(\phi_{22}+\phi_{22'}\right)$ and $\left(\phi_{21}+\phi_{21'}\right)$ at the liquid-liquid interface $\Sigma_{b2}$ as shown in figure~\ref{fig:Model} varying with angle $\varphi$ defined in the inset. (\textit{b}) Effect of the additional function $F(\varphi)$ with the same legend as in (\textit{a}). Examples are provided for increasing nondimensional distances $L/R_{d,0}=1.01, 1.10, 1.20, 1.40$ as marked with the black arrows in (\textit{a}) and (\textit{b}). (\textit{c}) Nondimensional radial velocity ${v_{R}}\,R_{d,0}^2/Q$ for $\rho_{o}/\rho_{w}=0.80$ and $L/R_{d,0}=1.40$ varying with the nondimensional radial distance $R/R_{d,0}$ for increasing angles $\varphi=0^{\circ}, 30^{\circ}, 60^{\circ}, 90^{\circ}$ as marked with the black arrow. (\textit{e}) Nondimensional tangential velocity ${v_{\varphi}}\,R_{d,0}^2/Q$  for $\rho_{o}/\rho_{w}=0.80$ and $L/R_{d,0}=1.40$ with the same legend as in (\textit{c}). The shaded areas in (\textit{c}) and (\textit{d}) denote the interior of the droplet ($R/R_{d,0}\leqslant\,1$). The calculations are valid for cases when the bubble does not contact the droplet.}
    \label{fig:model_verify}
\end{figure}

Here we verify the effect of the additional function $F(\varphi)$ using relations (\ref{eq:Frz2}) and (\ref{eq:BC22}). For relation (\ref{eq:Frz2}), according to table~\ref{Table1}, with given $L/R_{d,0}$, the ratio $-\left(\phi_{22}+\phi_{22'}\right)/\left(\phi_{21}+\phi_{21'}\right)$ is independent of time. Figure~\ref{fig:model_verify}(\textit{a}) displays the variation of the ratio with angle $\varphi=0^{\circ}$ -- $90^{\circ}$ for $L/R_{d,0}=1.01, 1.10, 1.20,$ and 1.40 on the interface $\upSigma_{b2}$. For $\varphi=0^{\circ}$, the ratio increases with increasing $L/R_{d,0}$, indicating that when the cavitation bubble is generated at a longer distance from the oil droplet, the line sources in the droplet play a more important role than the point sink. As the angle $\varphi$ increases from $0^{\circ}$ to $90^{\circ}$ (at the rigid boundary), the ratio increases monotonically to around 1, which indicates that the line sources have almost the same strength as the point sink at the rigid boundary. With the addition of $F(\varphi)$, as shown in figure~\ref{fig:model_verify}(\textit{b}), the ratio $|\left(\phi_{22}+\phi_{22'}+F\right)/\left(\phi_{21}+\phi_{21'}\right)|$ is about one to two magnitudes smaller than the ratio $|\left(\phi_{22}+\phi_{22'}\right)/\left(\phi_{21}+\phi_{21'}\right)|$. With increasing $L/R_{d,0}$, the ratio $|\left(\phi_{22}+\phi_{22'}+F\right)/\left(\phi_{21}+\phi_{21'}\right)|$ decreases fast to zero, e.g., within 1\% for $L/R_{d,0}>1.20$.  

On the other hand, for relation (\ref{eq:BC22}), we calculate the variations of the radial velocity $v_{R}$ and the tangential velocity $v_{\varphi}$ with $R/R_{d,0}=0$ -- 1.1, here with both velocities in the nondimensional forms. Examples are given for the kerosene droplet ($\rho_{o}/\rho_{w}$=0.80) at selected angles $\varphi=0^{\circ}, 30^{\circ}, 60^{\circ},$ and $90^{\circ}$, when a cavitation bubble is generated at $L/R_{d,0}=1.40$, as shown in figure~\ref{fig:model_verify}(\textit{c}) and~\ref{fig:model_verify}(\textit{d}). The calculated normal velocity is continuous at the droplet-water interface (figure~\ref{fig:model_verify}(\textit{c})), which agrees with the relation (\ref{eq:BC22}). By contrast, discontinuities occur to the tangential velocities at the interface (figure~\ref{fig:model_verify}(\textit{d})), indicating reasonable interfacial slippage.

\section{Contribution of the bubble-droplet interface to the anisotropy parameter at bubble collapse}\label{sec:app_theory}

\begin{figure}
	\centerline{\includegraphics[width=62 mm]{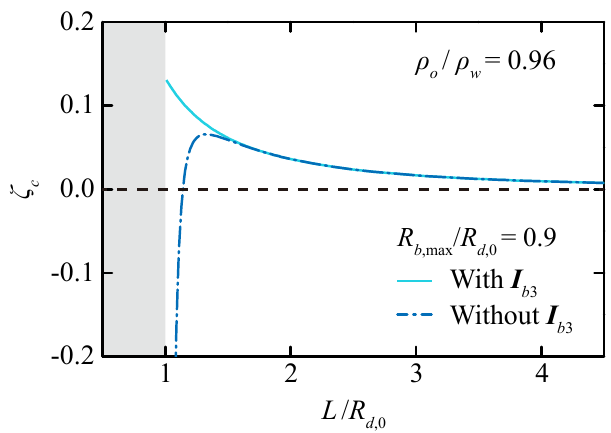}}
	\caption{\label{fig:ComparisonTheory} Comparison of the theory of the anisotropy parameter at the end of the bubble collapse $\zeta_{c}$ as a function of $L/R_{d,0}$ for $\rho_{o}/\rho_{w}=0.96$ and $R_{b,\max}/R_{d,0}=0.9$. The solid line represents the calculation with the contact term $\boldsymbol{I}_{b3}$, while the dash-dotted line represents the calculation without $\boldsymbol{I}_{b3}$. The shaded area denotes the interior of the droplet.}
\end{figure}

As has been mentioned in~\S\ref{sec:bubble_displacement}, with large $R_{b,\max}/R_{d,0}$, the bubble migrates at collapse towards the rigid boundary regardless of the dimensionless distance $L/R_{d,0}$. We provide in figure~\ref{fig:ComparisonTheory} the variation of $\zeta_c$ with $L/R_{d,0}$ when $R_{b,\max}/R_{d,0}$ is 0.9 for a silicone oil droplet. The solid line is the calculated curve with the component contribution of $\boldsymbol{I}_{b3}$ arising from the bubble-droplet interface $\upSigma_{b3}$, which approves of the invariability of the bubble migration direction. By contrast, the dash-dotted line shows the calculation result without $\boldsymbol{I}_{b3}$, which coincides with the solid line for $L/R_{d,0}\gtrsim\,1.5$ while it becomes negative at $L/R_{d,0}\lesssim\,1.2$. This indicates that the effect of bubble-droplet contact can be neglected for $L/R_{d,0}\gtrsim\,1.5$, although the bubble contacts the droplet at its maximum size within $L/R_{d,0}=1+R_{b,\max}/R_{d,0}=1.9$. On the other hand, ${I}_{b3}$ is positive, indicating that the bubble-droplet contact induces an attractive force on the bubble, which could arise from the rigid boundary immersed in the oil droplet. 

\bibliographystyle{jfm}
\bibliography{CavitationLib}

\end{document}